\documentclass[aps,prb,twocolumn,10pt,amsmath,amssymb,floatfix]{revtex4-2}
\usepackage{graphicx}
\graphicspath{{./}{../}}
\usepackage{bm}
\usepackage{mathtools}
\usepackage{ragged2e}
\usepackage{xcolor}
\usepackage{amsmath, amssymb}
\usepackage{float}   % put this in the preamble
\usepackage{subcaption}   % in preamble
\usepackage{hyperref}
\hypersetup{
    colorlinks=false,
    pdfborder={0 0 0}
}
\begin{document}

\title{Non-chiral edge states and cascading of exceptional points in the non-reciprocal Haldane model}

\author{Aditi A. Prabhudesai}
\thanks{These authors contributed equally to this work.}

\author{H. S. Chhabra}
\thanks{These authors contributed equally to this work.}

\author{Suraj S. Hegde}
 \email{surajhegde@iisertvm.ac.in}
\affiliation{ Indian Institute of Science Education and Research, Thiruvananthapuram \\
Kerala, India
}

\date{\today}

\begin{abstract}{Robust boundary-localised modes and band geometry arise in a variety of condensed matter lattice systems, of which the Haldane honeycomb model is a paradigmatic example. We study a non-Hermitian variant of the Haldane model, which exhibits two key features: distinctive exceptional-point (EP) structures in the periodic and open boundary spectra and non-chiral edge modes. In a fully periodic system, there are exceptional rings (ERs) that mark the boundary between the $\mathcal{PT}$-unbroken and $\mathcal{PT}$-broken regions, revealing Berry-curvature confinement across the $\mathcal{PT}$-symmetry transition. The system on a cylinder with zigzag boundaries exhibits a bifurcation transition of edge-to-bulk delocalisation.  We also observe a ``Russian-doll-like'' cascading of exceptional points as the non-Hermiticity is increased and we demonstrate how this is inherited from the bulk ER structure. The non-chiral edge modes remain robust against disorder in the non-Hermitian hopping parameter. The dynamical evolution shows a tunable range of weak non-Hermiticity and wave packet widths for which the edge-to-bulk delocalization time is extended. These results establish this model as a natural platform to explore Berry-curvature confinement, exceptional-point cascading and non-chiral edge physics. }
\end{abstract}

\maketitle
\section{Introduction}
Boundary localisation is a prominent phenomenon across a wide range of physical systems---both quantum and classical---and plays a central role in transport, response, and quantum computation. It is most commonly discussed in the context of topological phases, where boundary states exhibit distinctive features such as chiral propagation, fractional charge and statistics, and robustness against perturbations. In non-Hermitian, non-equilibrium, and open quantum systems, boundary physics is further enriched by the interplay between non-Hermitian skin effects, topological edge modes, and exceptional-point physics~\cite{Bardyn_2013,Ashida2021,Bergholtz2021,Ding2022,DeCarlo2022,Gneiting2022,Okuma2023,Meden2023,Sone_2026, yao-2018,Hegde-2023,Sone-2020,leykam-2017,zhu-2021,Zhu2022,cheng-2022,wang-2022}.
Non-Hermitian operators arise in diverse physical settings, including optical and photonic platforms, open quantum dynamics, disordered and interacting many-body systems, and active matter ~\cite{Bardyn_2013, Feng2017,Takasu2020,Ashida2021,Bergholtz2021,Song2021,Ding2022,DeCarlo2022,Gneiting2022,Lin2022,Okuma2023,Meden2023,Liu2023,Li2024,Sone_2026}. Non-Hermiticity fundamentally determines the spectral properties governing dissipative evolution and many-body Green's functions ~\cite{Savin2003,Zloshchastiev2014,Echeverri2018,Song2019,Altland2021,Michen2021,Purkayastha_2022,Lehmann2024}.

 Exceptional points (EPs), where eigenvalues and eigenvectors coalesce, are a defining feature of non-Hermitian systems~\cite{Heiss2012}. These degeneracies can form exceptional rings (ERs) which are compact loci of EPs that exhibit nontrivial topology, unconventional boundary states~\cite{Xu-01-2017,Carlstrom2018,Cerjan2018,okugawa-2019,Budich-01-2019,Moors-01-2019,Wang-02-2019,Yoshida-03-2019,yoshida-04-2019,wang-10-2019,Liu-11-2019,Rui-12-2019,isobe-2021,staalhammar2021,isobe-2023,lei-2025,huang-2025,Qin2025}, and nonlinear Hall responses via the Berry-curvature dipole~\cite{Qin2025}. Related ER structures also arise in the Majorana-basis Haldane model of disordered Kitaev spin liquids~\cite{bergholtz-2022,fukui-2024}, which hosts non-Abelian spin liquid phases and chiral Majorana modes.  This motivates our study of EP physics and edge phenomenology in a non-Hermitian variant of the Haldane model with boundaries. We uncover three remarkable aspects of this exceptional phase -- 1) The normalised right-state Berry curvature is confined within the ERs with the same sign structure at both valleys, revealing distinct eigenstate geometry across the $\mathcal{PT}$-symmetry transition. 2) The system hosts unusual disorder-robust non-chiral edge states whose dynamical localization can be controlled through the wave-packet width and the non-Hermitian parameter. 3) The edge states join the bulk through an EP and as the non-Hermiticity increases, there is a cascade of EPs forming a `Russian-doll' pattern in the spectrum.

\begin{figure}[t]
    \centering\normalcolor
    \includegraphics[width=7cm, height=4cm]{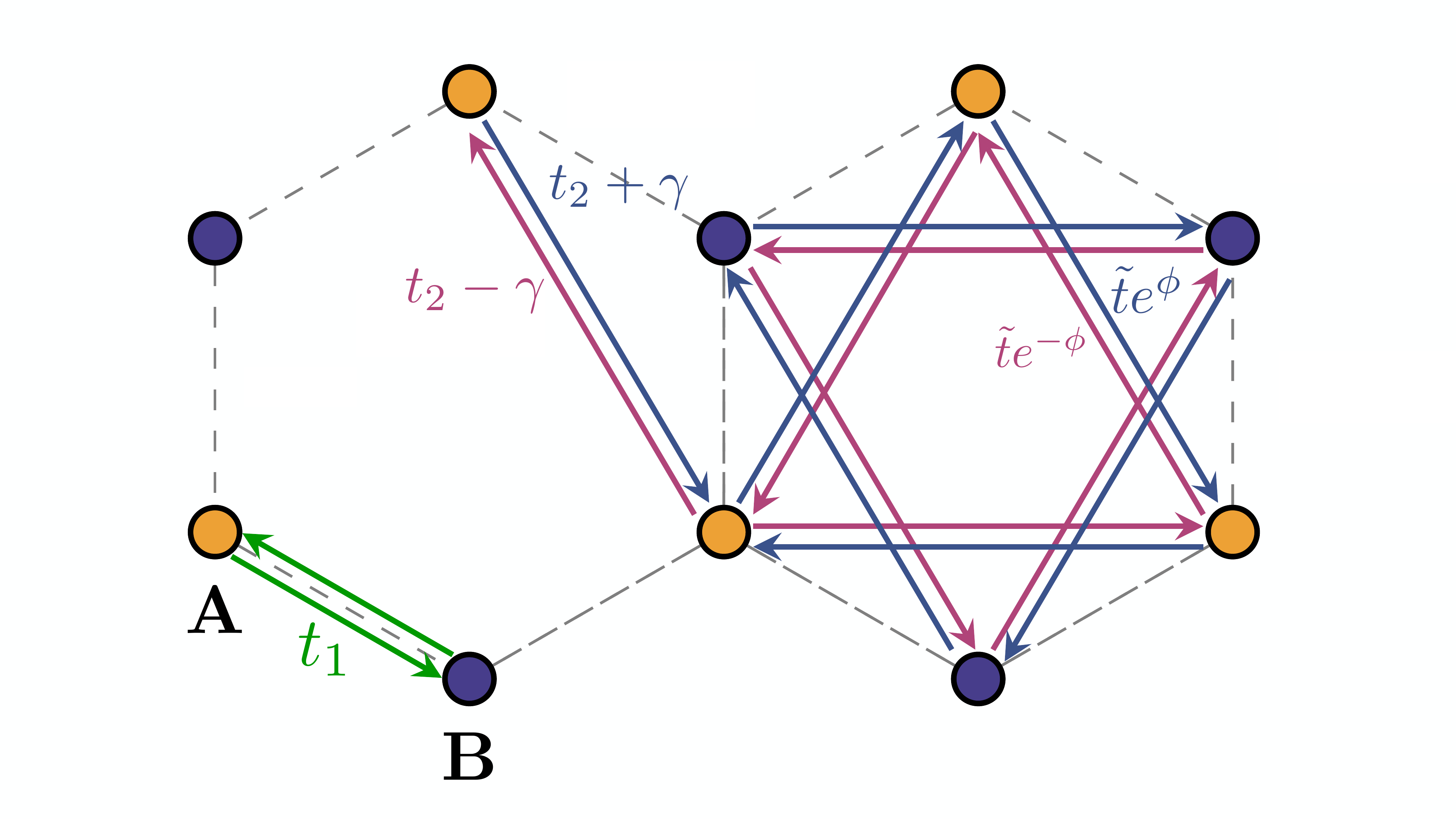}
    \caption{ \justifying Honeycomb lattice with sublattices A and B. The dashed lines and colored arrows indicate nearest-neighbor (NN) and next-nearest-neighbor (NNN) hoppings respectively. The non-reciprocal NNN hopping can equivalently be interpreted as hopping in the presence of an imaginary flux.}
    \label{fig:lattice}
\end{figure}
\begin{figure*}[t]
\centering\normalcolor

\begin{minipage}[t]{0.32\textwidth}
    \centering
    \includegraphics[width=\linewidth,height=4.5cm,keepaspectratio]{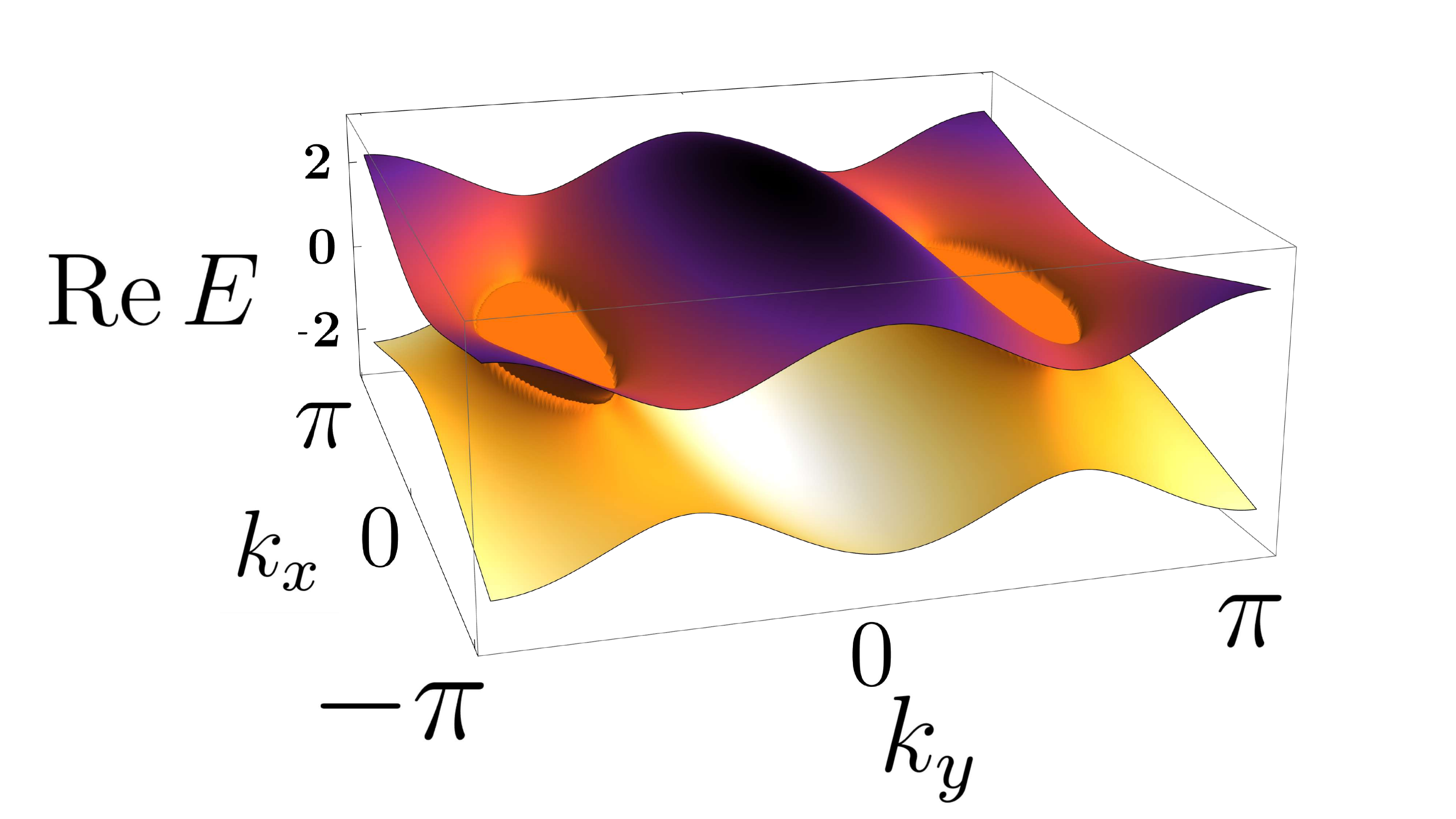}
    \\ (a)
\end{minipage}
\hfill
\begin{minipage}[t]{0.32\textwidth}
    \centering
    \includegraphics[width=\linewidth,height=4.5cm,keepaspectratio]{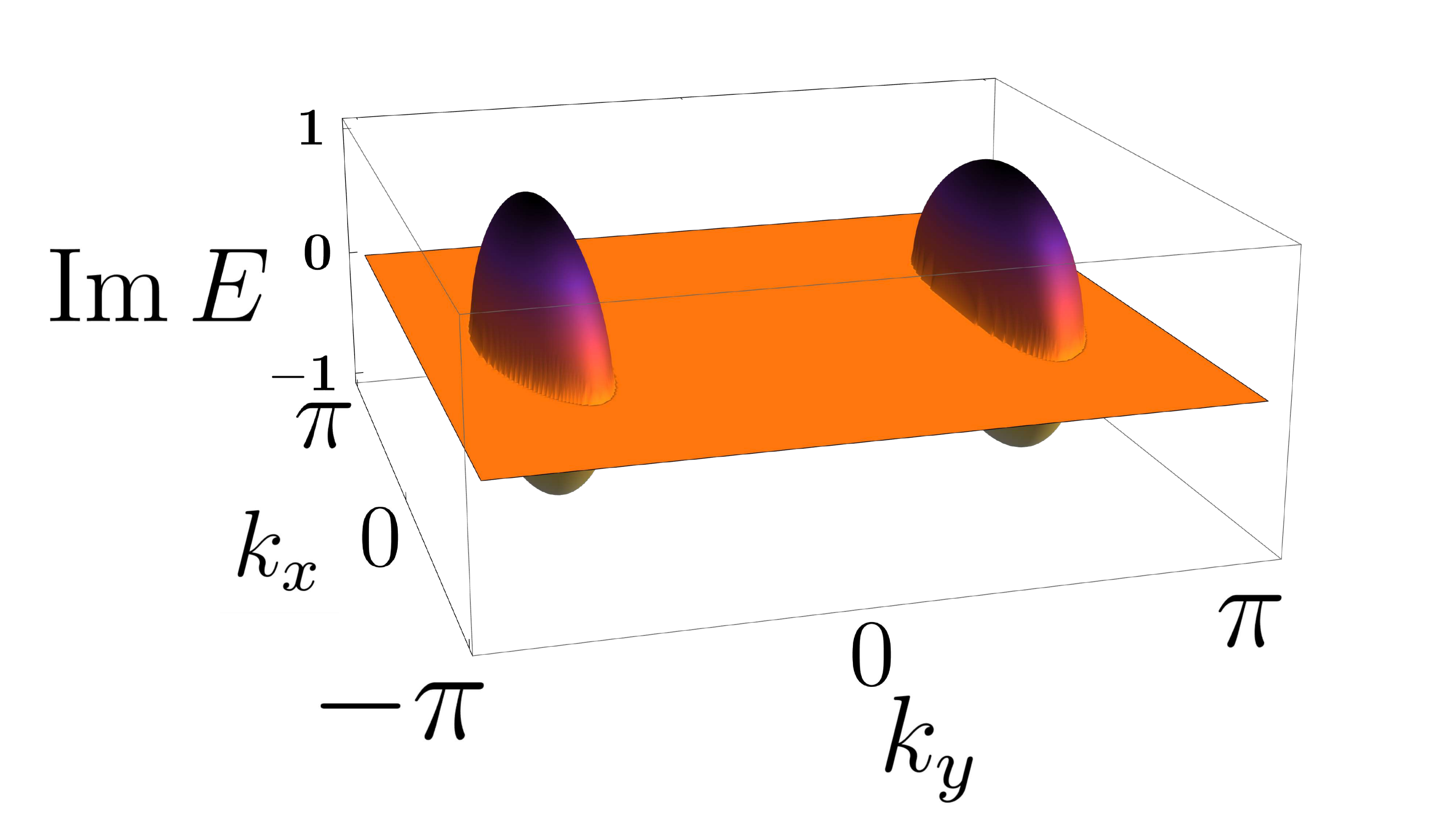}
    \\ (b)
\end{minipage}
\hfill
\begin{minipage}[t]{0.32\textwidth}
    \centering
    \includegraphics[width=\linewidth,height=4.5cm,keepaspectratio]{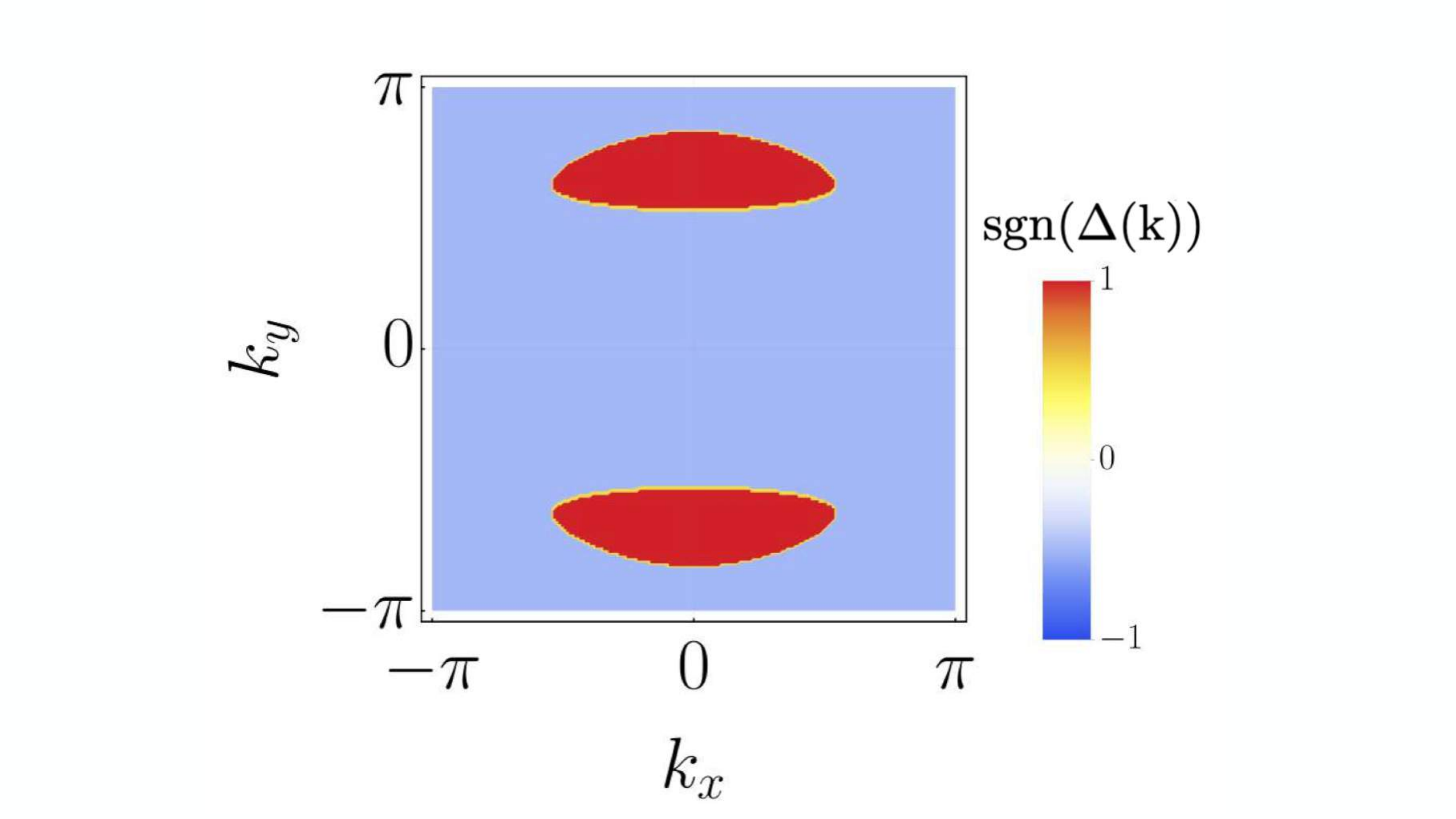}
    \\ (c)
\end{minipage}

\caption{\justifying
(a),(b) Real and imaginary parts of the complex energy spectrum for the Haldane model with non-reciprocal NNN hoppings under periodic boundary conditions.
(c) The local $\mathbb{Z}_2$ invariant, {$\mathrm{sgn}(\Delta(\mathbf{k}))$}, plotted over the Brillouin zone.
This invariant distinguishes the $\mathcal{PT}$-symmetry-unbroken and $\mathcal{PT}$-symmetry-broken regions of the spectrum.
Exceptional rings mark the boundaries where the $\mathcal{PT}$-symmetry-breaking transition occurs. (Parameters: $t_1 = 1$, $\tilde{t} = 0.22$, $\phi = 0.8$)}
\label{fig:spectrum-panels}
\end{figure*}
\section{The Model} 

We consider a tight-binding Hamiltonian on a two-dimensional honeycomb lattice with both nearest-neighbour (NN) hopping and next-nearest-neighbour (NNN) hoppings~\cite{haldane-1988}. The phase in the NNN hoppings of the Haldane model is interpreted as an effective hopping amplitude in the presence of a magnetic flux (though the net flux through the system is zero). We consider non-reciprocity in NNN hopping that can be rewritten effectively as hopping in the presence of an imaginary flux {~\cite{imag-gauge-field-1,imag-gauge-field-2}} (Fig.~\ref{fig:lattice}). The resulting Hamiltonian for the `imaginary-flux' Haldane model is then given by

\begin{displaymath}
\begin{aligned}
\mathcal{H}_{\mathrm{NH\text{-}Haldane}}
&=  t_1 \sum_{\langle i,j\rangle} c_i^\dagger c_j \\
&\quad
+ \sum_{\langle\!\langle i,j\rangle\!\rangle }
   \tilde{t}\!\left(
     e^{-\phi }\, c_i^\dagger c_j
   + e^{\phi }\, c_j^\dagger c_i
   \right)
\end{aligned}
\end{displaymath}
where $\langle i,j\rangle$($\langle\!\langle i,j\rangle\!\rangle$) and $t_1$ ($\tilde{t}$)
denote NN (NNN) site pairs and their hopping amplitudes, respectively. The NN sum includes both directions; each NNN bond is counted once with the oriented Haldane convention. The exact mapping between the non-reciprocal hopping $\gamma$ and hopping under `imaginary-flux' is as follows: $\tilde{t} e^{\pm \phi} = t_2 \pm \gamma $, valid for $|\gamma|<t_2$,  where $t_2$ is the usual NNN hopping amplitude, $\tilde{t} = \sqrt{t_2^{\,2}-\gamma^{\,2}}$ and {$\phi = \frac{1}{2}\ln\!\left(\frac{t_2+\gamma}{t_2-\gamma}\right)$}. Note that one can go beyond this mapping between non-reciprocal hopping and imaginary flux, and in general treat $\tilde{t}$ and $\phi$ as independent variables keeping  $\tilde{t}e^{|\phi|}<t_1$. We also note that the signs of the imaginary flux on the NNN links are such that the directions of non-reciprocity are opposite for the A and B sublattices, similar to the flux-sign convention found in the Hermitian Haldane model. This is crucial for generating the imaginary Haldane mass that gives rise to all the exceptional point physics. Related non-Hermitian Haldane models, including those with non-reciprocal NNN terms, have been explored previously~\cite{Ezawa2019,rezendiz-2020,Zhu2022,WangYi2023,Wei2023,Jiang2024,Xie2025,dong-2025,Chen2025}. The real asymmetric NNN hoppings considered here are contained in the broader non-Hermitian Haldane-model family studied in Ref.~\cite{dong-2025}, which also identified exceptional rings around both valleys. Our focus is on the right-eigenstate geometry, the nested exceptional-point cascade under transverse confinement, and the non-chiral boundary dynamics of the imaginary-flux regime. 

\begin{figure}[!t]
\centering\normalcolor
    \includegraphics[width=5.2cm]{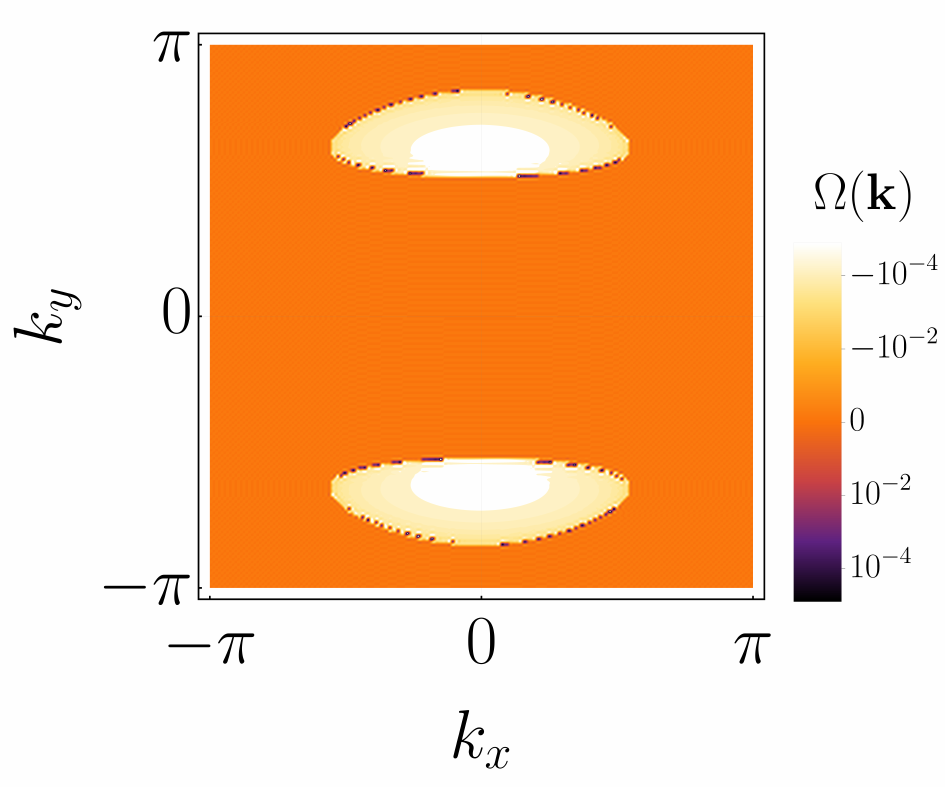}
    \caption{\justifying
    Berry-curvature landscape of exceptional rings (ERs) in the imaginary-flux Haldane model. The ERs act as Berry-curvature flux tubes, leading to strong localization of the Berry curvature within their interior and particularly in this model, the Berry curvature has the same sign in both the ERs around $K$ and $K'$ points of the Brillouin zone. Parameters: $t_1=1$, $\tilde{t}=0.22$,  $\phi=0.8$}
    \label{fig:berry-imag-flux}
\end{figure}
\begin{figure*}[t]
    \centering\normalcolor
    \includegraphics[width=\linewidth,keepaspectratio]{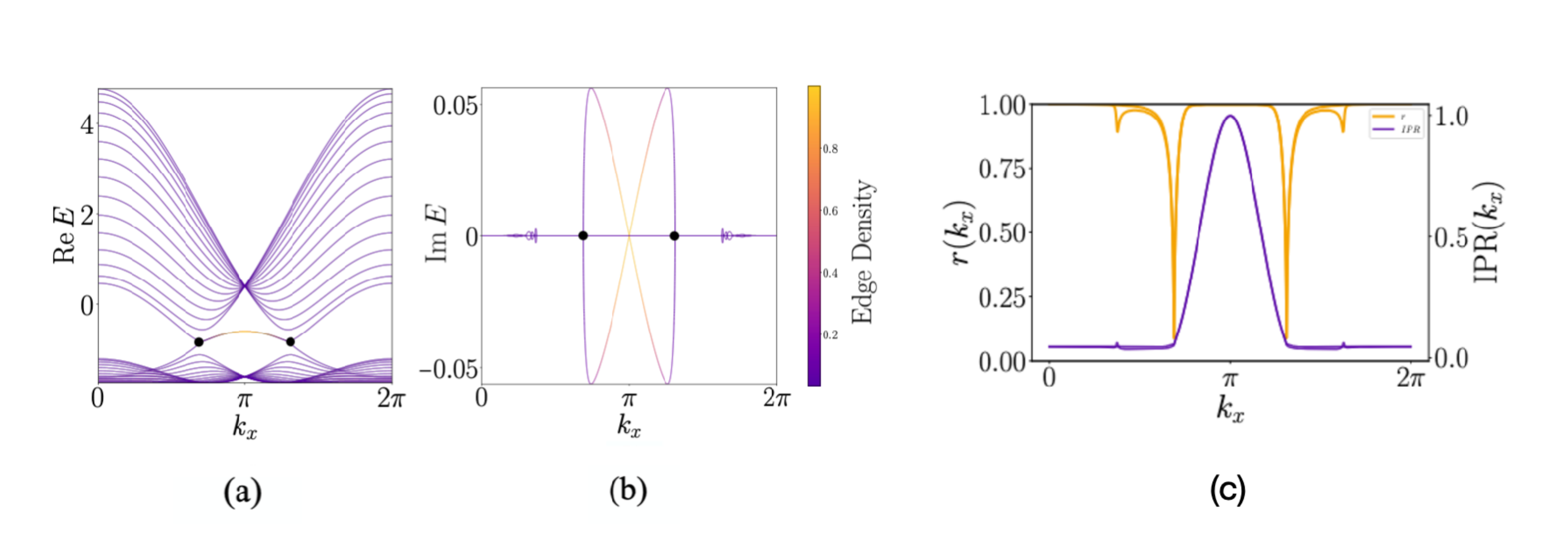}
    \caption{\justifying{(a),(b) Real and imaginary parts of the spectrum of the imaginary-flux Haldane model on a cylinder with zigzag boundaries. Edge-localised states are highlighted via the probability density at the left edge. Nonchiral edge states occur at $k_x=\pi$ with
    $\partial_{k_x}(\mathrm{Re}\,E)=0$ and finite
    $\partial_{k_x}(\mathrm{Im}\,E)$. They undergo a bifurcation transition at the exceptional points (black dots) and merge with the bulk. (c) Phase rigidity and inverse participation ratio (IPR) of the edge states as a function of $k_x$. The phase rigidity vanishes at the exceptional points, accompanied by suppression of the IPR, signalling delocalisation of the edge states into the bulk. (Parameters: $N_y = 30$, $t_1 = 1$, $\tilde{t} = 0.3$, $\phi = 0.05$.)}}
    \label{fig:spectrum-pr-ipr}
\end{figure*}

Using periodic boundary conditions along both spatial directions, the Fourier-transformed Hamiltonian is expressed as $H(\vec{k})=\sum_{i=0}^{3}h_i(\vec{k})\cdot\sigma_i$ where
$h_0(\vec{k})=
2\tilde{t}\cosh\phi\,\big[\cos(\theta_1)+\cos(\theta_2)+\cos(\theta_1-\theta_2)\big]$,
$h_1(\vec{k})=t_1\big[1+\cos(\theta_1)+\cos(\theta_2)\big]$, $h_2(\vec{k})=t_1\big[\sin(\theta_1)+\sin(\theta_2)\big]$, and $h_3(\vec{k})=2 i\tilde{t}\sinh\phi\,\big[\sin(\theta_1)-
\sin(\theta_2)-\sin(\theta_1-\theta_2)\big]$
with lattice vectors
\( \vec{a}_1 = \frac{a}{2}(3,\sqrt{3}) \),
\( \vec{a}_2 = \frac{a}{2}(3,-\sqrt{3}) \),
 \( a=1 \) and $\theta_i=\vec{k}\cdot \vec{a}_i$. The imaginary-flux term acts as an imaginary Haldane mass added to the graphene Hamiltonian. The resulting Bloch Hamiltonian satisfies spinless time-reversal symmetry, $H^{*}(\mathbf{k}) = H(-\mathbf{k}),$ while it lacks chiral symmetry $\Gamma H^{\dagger}(\vec{k})\Gamma^{-1}=-H(\vec{k})$ placing it, with crystalline symmetries set aside, in the non-Hermitian Altland--Zirnbauer symmetry class AI~\cite{gong-2018}.

% Allowing the NNN flux to be in general complex yields a mass prefactor $2\tilde{t}\sin(\Phi+i\phi)$. In this complex-flux plane, the Hermitian Haldane model lies on the real axis, while our model occupies the imaginary axis. A generic complex-flux extension breaks TRS and supports gapped edge states with finite lifetimes~\cite{dong-2025,SM}. We instead focus on the imaginary-flux limit which is a singular exceptional phase hosting $\mathcal{PT}$-protected ERs.

We observe the emergence of a pair of ERs, i.e. closed loci of EPs in the Brillouin zone (BZ) (Fig.~\ref{fig:spectrum-panels}(a),(b)). The ERs are seen to form around the special points $K,K'$ in the BZ of the hexagonal lattice. It is typically seen that the addition of non-Hermiticity to models having degeneracies or avoided crossings leads to the formation of EPs ~\cite{Heiss_1990,Heiss2012}. In our context, the addition of a time-reversal-preserving imaginary Haldane term turns the gapless points of graphene at $K,K'$ into ERs.
Parity $P=\sigma_x$ combines with time reversal to give the symmetry $PH^*(\mathbf k)P^{-1}=H(\mathbf k)$. Writing $h_3=ig$ with real $g$, the spectrum is
$E_\pm=h_0\pm\sqrt{\Delta}, \Delta=h_1^2+h_2^2-g^2.$ The region enclosed by the ERs corresponds to the $\mathcal{PT}$-symmetry broken phase with complex eigenvalues, while $\mathcal{PT}$ symmetry remains unbroken elsewhere in the BZ~\cite{bender-boettcher}. We evaluate a $\mathbb{Z}_2$ invariant,
$\mathrm{sgn}\!\left[\Delta(\mathbf{k})\right]$, throughout the Brillouin zone, as shown in Fig.~\ref{fig:spectrum-panels}(c). The invariant distinguishes the $\mathcal{PT}$-unbroken ($\Delta>0$) and $\mathcal{PT}$-broken ($\Delta<0$) regions and changes sign precisely at $\Delta=0$, thereby identifying the exceptional rings as the boundaries between the two phases. The ERs thus delineate the transition between these regions, characterised by the local $\mathbb{Z}_2$ invariant {$\mathrm{sgn}[\Delta(\mathbf{k})]$} ~\cite{gong-2018,okugawa-2019}.

A notable feature in such $\mathcal{PT}$-symmetry breaking ERs is the confinement of the normalised right-state Berry curvature inside the ERs (Fig.~\ref{fig:berry-imag-flux}): 
$\boldsymbol{\Omega}_n=\nabla_{\mathbf{k}}\times\mathbf{A}_n$ and
$\mathbf{A}_n=-i\langle u^R_n|\nabla_{\mathbf{k}}u^R_n\rangle$.
For a general Hamiltonian of the form $H=h_0\sigma_0+h_x\sigma_x+h_y\sigma_y+ih_z\sigma_z$, with real coefficients in a fixed basis and $\mathcal{PT}=\sigma_xK$, setting $\epsilon=E-h_0$, with $h_x-ih_y\ne0$, the normalised right eigenstate is
$|u_R\rangle=(1,\rho e^{i\chi})^T/\sqrt{1+\rho^2}$, where
$\rho e^{i\chi}=(\epsilon-ih_z)/(h_x-ih_y)$.
This gives
$\Omega=2\rho(1+\rho^2)^{-2}(\nabla\rho\times\nabla\chi)_z$.
Outside an ER, $\epsilon$ is real and $|\epsilon-ih_z|^2=h_x^2+h_y^2$, hence $\rho=1$ and $\Omega=0$.
Inside the ER, $\epsilon=i\lambda$ with real $\lambda$, so that $\rho$ can vary and permit a nonzero curvature.

This establishes the confinement of the right-state Berry curvature inside the ER itself. Hence the ERs act as Berry Curvature Flux (BCF) tubes in sharp contrast to the distributed Berry curvature of, e.g., the complex-flux Haldane model (See Appendix ~\ref{app:complexflux}). In the complex flux regime, with flux $\Phi+i\phi$, the mass prefactor becomes $2\tilde{t}\sin(\Phi+i\phi)=2\tilde{t}\,(\sin\Phi\cosh\phi+i\cos\Phi\sinh\phi)$. The conventional Haldane model sits on the real axis, $\phi=0, \Phi= \pi/2$, and our model on the imaginary axis, $\Phi=0$. Away from the imaginary axis, both time-reversal and $\mathcal{PT}$ symmetries are broken and the edge states are gapped with finite lifetimes~\cite{dong-2025}.
A similar BCF-tube structure arises when the Haldane term is replaced by an imaginary
Semenoff mass $\pm iM$ with real $M$, though there the two ERs carry \emph{opposite} signs of Berry
curvature, unlike the imaginary-flux case (See Fig. \ref{fig:complex-flux-all}(d)). Such localised Berry curvature and ERs may have non-trivial consequences in the transport properties of the model.

\section{The model on a cylinder with zigzag boundaries}

We next consider the system in a cylindrical geometry with open zigzag boundaries along one direction (taken to be the $y$-direction), while retaining periodicity along the $x$-direction. In this configuration, the momentum along the $x$-direction is conserved, and the Hamiltonian reduces to an effective one-dimensional $H(k_x)$. It can be written as:

 \begin{equation}
\begin{aligned}
H(k_x) =\;&
\sum_{i=1}^{N_y} \psi_i^\dagger\, h_{i,i}(k_x)\,\psi_i +\sum_{i=1}^{N_y-1}
\Big[
\psi_{i+1}^\dagger\, h_{i+1,i}(k_x)\, \psi_i\\
&
+ \psi_i^\dagger\, h_{i,i+1}(k_x)\, \psi_{i+1}
\Big].
\end{aligned}
\label{eq:obc-hamiltonian}
\end{equation}
The intra-layer block is given by $h_{i,i}(k_x)=2\tilde t\cos k_x\cosh\phi\,\mathbb I
-2i\tilde t\sin k_x\sinh\phi\,\sigma_z
+2t_1\cos(k_x/2)\sigma_x$ ,
while the inter-cell hopping block is $h_{i,i+1}(k_x)=2\tilde t\cos(k_x/2)\cosh\phi\,\mathbb I
+2i\tilde t\sin(k_x/2)\sinh\phi\,\sigma_z
+t_1\sigma^+$, and $h_{i,i-1}(k_x)=2\tilde t\cos(k_x/2)\cosh\phi\,\mathbb I
+2i\tilde t\sin(k_x/2)\sinh\phi\,\sigma_z
+t_1\sigma^-$.
Here, {$k_x$} is the momentum along the $x$ direction. Note that we have switched from $k_x \in [-\pi,\pi]$ to $k_x \in [0,2\pi]$ for better visual representation of the EPs in the plots. 

% Next we consider the model under open boundary conditions along one edge and periodic boundary conditions in the other, effectively putting it on a cylinder such that it exhibits zigzag edges. We focus on zigzag boundaries, which in our model give rise to edge states and EPs. 

We first consider the system with a small imaginary flux $\phi$ introduced through non-reciprocal NNN hopping $\gamma\ll t_2$. This system exhibits a complex spectrum as shown in Figs.~\ref{fig:spectrum-pr-ipr}(a),(b) colored by left-edge density. The maximally localised edge state occurs at $k_x=\pi$.
Near $k_x=\pi$, a smooth edge branch has the expansion
$E_{\mathrm{edge}}(\pi+\delta k)=E_0+i v_I\delta k+\tfrac12 a_R\delta k^2+\mathcal{O}(\delta k^3)$,
where $E_0=E_{\mathrm{edge}}(\pi)$, $v_I=\partial_{k_x}\mathrm{Im}E_{\mathrm{edge}}|_\pi$ and $a_R=\partial_{k_x}^2\mathrm{Re}E_{\mathrm{edge}}|_\pi$ are determined by the boundary spectrum. This form follows from $E_{\mathrm{edge}}(\pi+\delta k)=E_{\mathrm{edge}}^*(\pi-\delta k)$ for a branch mapped to itself by time reversal.
\begin{figure}[t]
    \centering\normalcolor
    \includegraphics[width=0.9\linewidth, trim=8cm 5cm 0 0, clip, keepaspectratio]{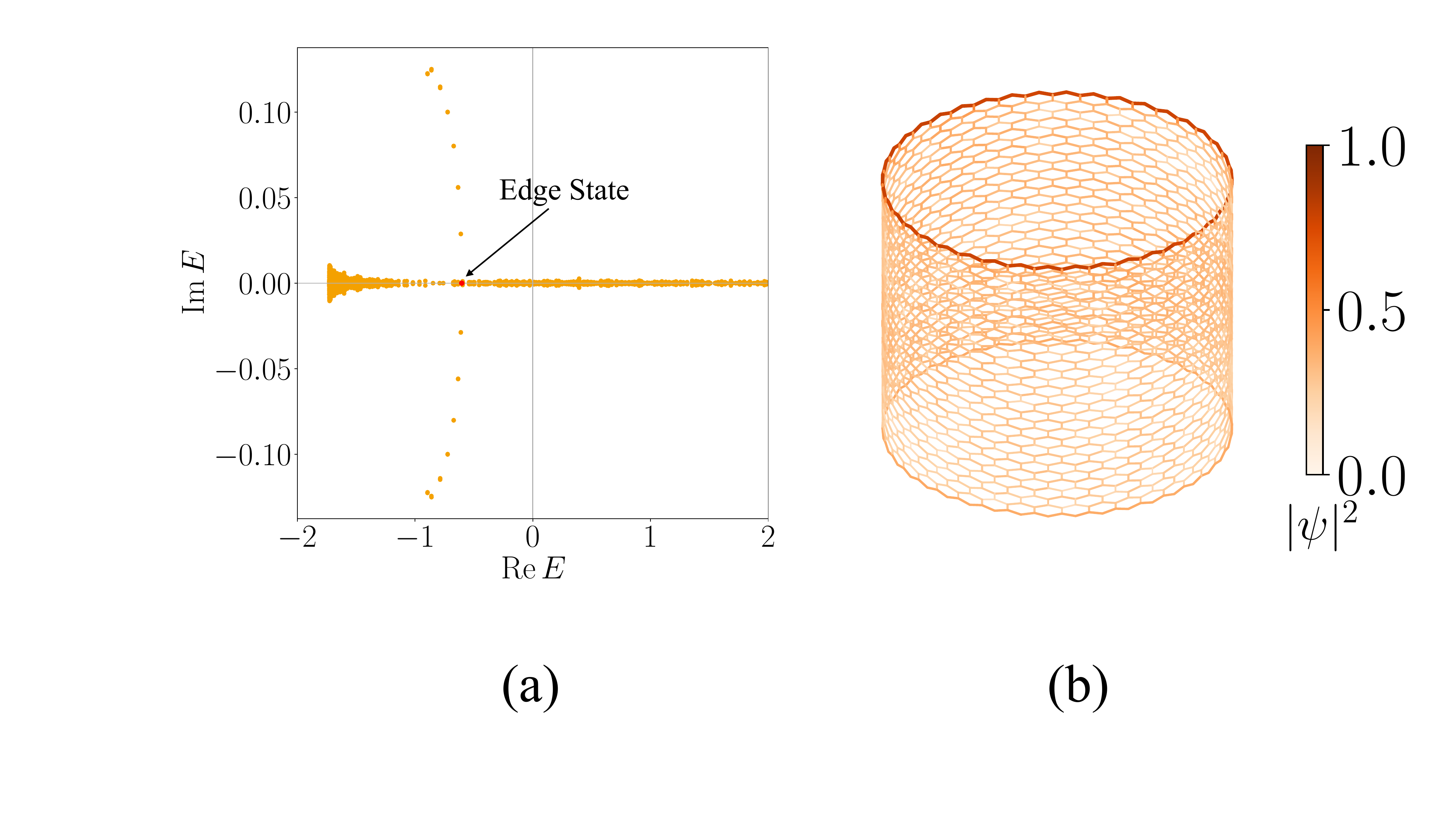}
    \caption{\justifying{Robust non-chiral edge states along zigzag boundaries of the imaginary-flux Haldane model with disorder in the NNN hoppings (imaginary flux). (a) In the complex plane of eigenvalues, the edge states form a branch projecting out along the imaginary axis. (b) The state at $\mathrm{Im}\,E =0$, $\mathrm{Re}E \sim -2\tilde{t}$ (marked red in (a)) remains pinned to those values and localised at the edge even at large disorder strength $\phi_m \sim \phi_D,$ thus demonstrating robustness. (Parameters: $N_x = 40$, $N_y = 30$, $t_1 = 1.0$, $\tilde{t} = 0.3$. Disordered values taken from a distribution with mean $\phi_m = 0.1$ and window width $\phi_D=0.1$.)}}
    \label{fig:nanotube}
\end{figure}
As opposed to the Haldane model, the real part of the dispersion of the edge states shows no chirality at $k_x=\pi$, i.e., $\text{Re}(\partial E/ \partial k_x) \sim 0$  and has very small dispersion even away from it. On the other hand, the imaginary part of the edge-state spectrum is linear in $k_x$ close to $k_x=\pi$. The ``chirality'' in the imaginary part has consequences in the wave packet dynamics as we will see in Section \ref{sec:wavepacket}. 

In this open-boundary configuration, we observe the emergence of the EPs in the edge-state spectrum. These EPs are identified through the phase rigidity, $r = \frac{|\langle \psi_L | \psi_R \rangle|}{\sqrt{\langle \psi_L | \psi_L \rangle \langle \psi_R | \psi_R \rangle}}$~\cite{wiersig-2023,brody-2013}, which quantifies the biorthogonality between the left and right eigenstates. As shown in Fig.~\ref{fig:spectrum-pr-ipr}(c), $r$ approaches zero precisely at the EPs, providing a clear spectral signature of eigenstate coalescence. 

To further characterise the spatial structure of these states, we compute the inverse participation ratio (IPR) of the right eigenstate $|\psi_R\rangle$, defined as $\mathrm{IPR} = \frac{\sum_i |\psi_{R,i}|^{4}}{\left(\sum_i |\psi_{R,i}|^{2}\right)^{2}}$. The IPR provides a quantitative measure of spatial localization, as it remains finite for localised states and decreases toward zero for delocalised states, scaling as $O(N_y^{-1})$ for extended bulk states. We find that it decreases to a finite-width bulk scale $O(N_y^{-1})$ near the EPs as shown in Fig.~\ref{fig:spectrum-pr-ipr}(c), indicating that the edge states delocalise into the bulk in the vicinity of these points, marking a bifurcation transition.

The edge states merge into the bulk near the EPs. The transfer-matrix condition $|\lambda|=1$ marks loss of exponential localization in the semi-infinite limit (See Appendix \ref{app:main-model}). Because the bulk bands meet at EPs, these edge states are not characterised by the usual isolated-band Chern number. At $k_x=\pi$, the effective one-dimensional model has $|\nu|=1$ for $4|\tilde t\sinh\phi|<|t_1|$, as shown in the Appendix \ref{app:main-model}. The gap closes at equality. Away from $k_x=\pi$, this winding construction does not provide an integer-valued invariant for each slice. Alternative characterizations have been used in the case of symmetry-broken extensions of the SSH model~\cite{Jiang2018,Hattori_2024,Zhong2025}.

Finally, we numerically study the real-space lattice Hamiltonian on a cylinder with disorder in the non-reciprocity. The spectrum and the edge density profile are shown in Fig.~\ref{fig:nanotube}. Even in the presence of disorder, we see that the mode that corresponds to $k_x=\pi$  in the clean case remains near its clean-limit energy and is robustly localised at the zigzag boundaries. The entire set of states that branch out along the imaginary axis also demonstrates robust localisation. The edge states vanish when the average non-reciprocity becomes comparable to the nearest-neighbour hopping strength or at very high disorder. This demonstrates persistence of the non-chiral edge modes within the hopping-disorder ensemble and parameter range studied.

\section{Cascading of exceptional points}

\begin{figure}[!t]
    \centering\normalcolor
    \includegraphics[width=\linewidth]{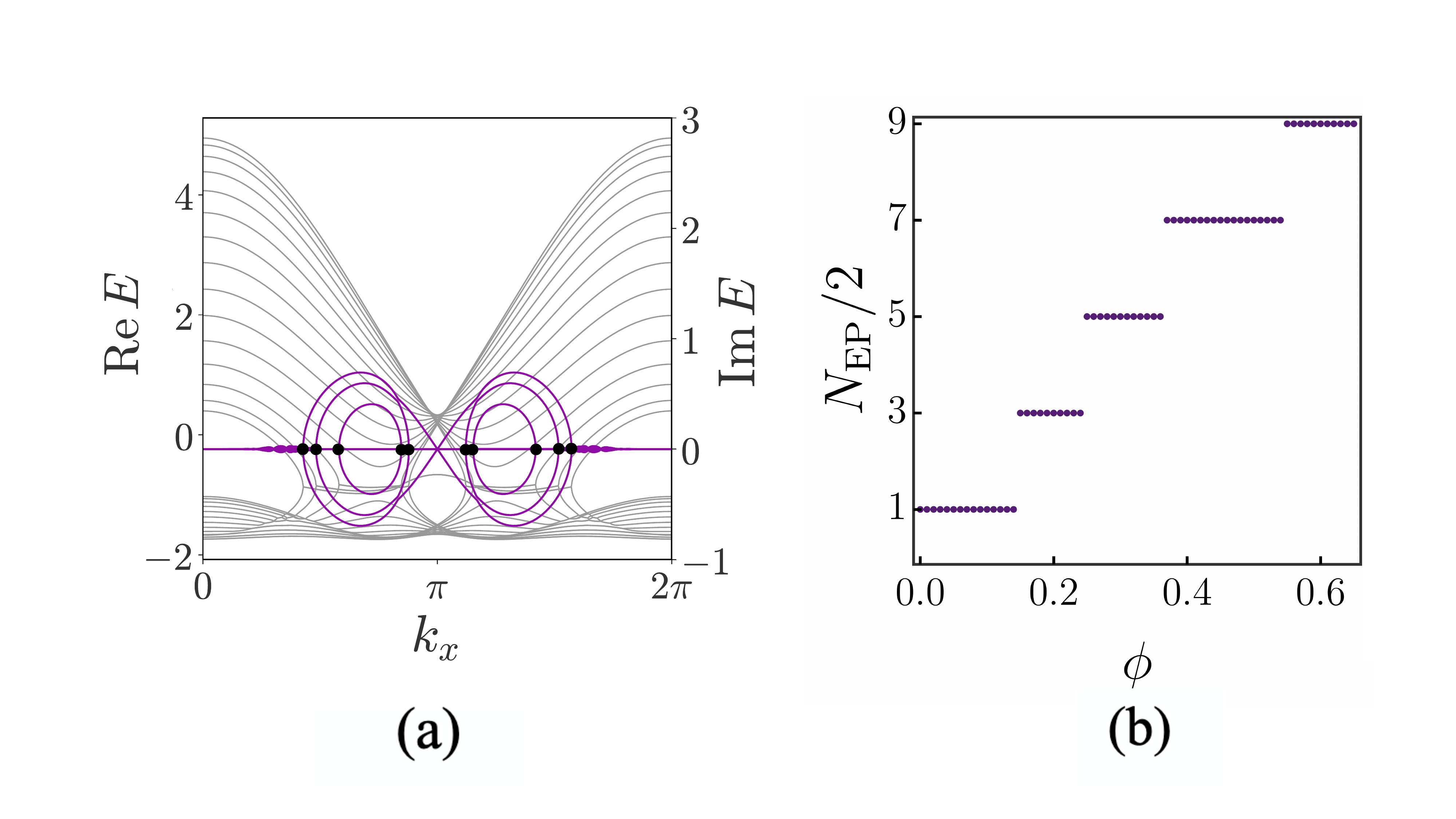}
   \caption{\justifying
    (a) Imaginary part of the bulk spectrum overlaid on the real part at larger flux ($\phi = 0.45$) for $N_y = 30$ and $\tilde{t} = 0.3$. Multiple EP pairs (marked with black dots only in the imaginary spectrum) are organised into a nested Matryoshka-like structure arising from successive bulk band coalescences.
    (b) Step-like structure in proliferation of EP pairs as a function of the non-Hermiticity parameter $\phi$ for $N_y = 40$ and $\tilde{t} = 0.3$.}
    \label{fig:edge-exceptional-band}
\end{figure}

We now turn to the high-flux regime. As the flux $\phi$ is increased, the number of EPs in a window of real energy evolves in a step-like manner as seen in Fig.~\ref{fig:edge-exceptional-band}(b). The first pair of EPs at the edge-bulk bifurcation persists for a certain interval in  flux $\phi$, defining the initial plateau. At a larger value of flux, bulk states lying adjacent to the edge-state energies coalesce, forming two additional pairs of EPs. Further increase in $\phi$  repeats this mechanism for successive nearby bulk states leading to discrete jumps in the total number of EP pairs.  All the bulk EPs remain confined within the region bounded by the edge-EPs, forming a nested Matryoshka-like (Russian-doll) structure visible in the imaginary spectrum Fig.~\ref{fig:edge-exceptional-band}(a).

The cascade mechanism and approximate bulk EP locations follow from the ERs of the fully periodic system through a bulk-spectrum zone-folding construction. For a zigzag cylinder, the open transverse boundary replaces the continuous two-dimensional bulk spectrum by a discrete set of bulk-like modes labelled by $q_{\perp,n}$, while $k_x$ remains conserved. In this approximation, each mode is described by an effective two-band Hamiltonian $H_n(k_x)=d_0\sigma_0+d_x\sigma_x+d_y\sigma_y+i\Gamma\sigma_z$ and develops exceptional points when $\Delta_n=d_x^2+d_y^2-\Gamma^2=0$. As the non-Hermiticity $\phi$ increases, the exceptional rings of the fully periodic system expand and successively encompass additional allowed transverse modes. Each newly admitted mode produces further bulk EP pairs, giving the stepwise cascade observed in the finite-width spectrum. Including the initial edge-bulk EP pair and the two symmetry-related valleys, the observed regime is summarised schematically by $N_{\mathrm{EP}}/2=1+2\sum_n\Theta(\phi-\phi_{c,n})$, which accounts for the odd plateaus $1,3,5,\ldots$ (See Appendix \ref{app:SM-zone-folding}).
Near either Dirac point, modes closer to its transverse momentum produce outer EP pairs, while more distant modes form smaller inner pairs, explaining the Matryoshka-like nesting. For a fixed low-lying band counted from a valley and for weak non-Hermiticity, $\phi_{c,n}\sim N_y^{-1}$, which implies that wider cylinders develop bulk EPs at lower non-reciprocity. The `arc' of bulk states between the EPs contains states with the largest imaginary parts corresponding to that band. At late times, normalised evolution selects such largest accessible $\mathrm{Im}E$ among initially populated states. 

\section{Wave packet dynamics of the edge band}
\label{sec:wavepacket}
\begin{figure}[!t]\centering\normalcolor\includegraphics[width=\linewidth]{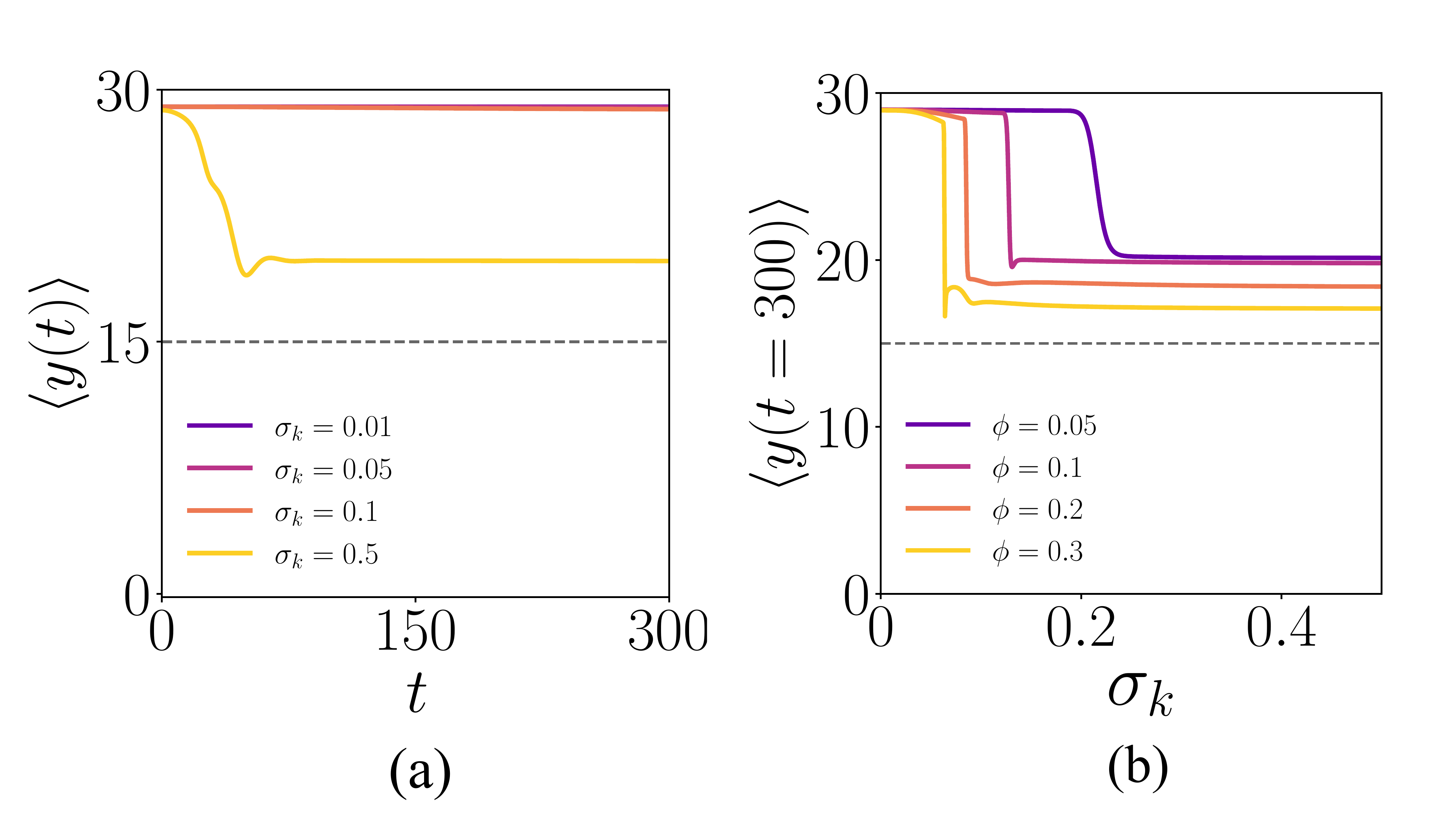}
    \caption{\justifying
    (a) Dynamics of $\langle y(t)\rangle$ showing finite-time edge persistence for weak non-Hermiticity ($\phi=0.05$) over a range of wave packet widths, with edge-to-bulk redistribution at a larger width due to self-acceleration.
    (b) Width-tolerance window for finite-time edge persistence narrows with increasing non-Hermiticity and widens as it decreases.}
    \label{fig:wavepacket-dynamics-edge}
\end{figure}
We further study the wave packet dynamics to examine how non-Hermitian spectral properties manifest in steady-state behavior, particularly with respect to edge modes. We construct a wave packet from a subset of bands, weighted by a Gaussian
$w(k_x)=\frac{1}{\sqrt{2\pi\sigma_k^2}}\exp[-(k_x-k_0)^2/(2\sigma_k^2)]$.
The time-evolved wave-packet is given by
$\Psi(y,t)=\sum_m \int dk_x\, e^{-iE_m(k_x)t}\, w(k_x)\, |\psi_{R,m}(k_x,y)\rangle$,
where $|\psi_{R,m}(k_x,y)\rangle$ is the $m$-th right eigenvector.  Momentum-space dynamics are tracked via $W(k_x,t)=|\langle \psi_{L,n}(k_x)|\Psi(y,t)\rangle|^2
=|e^{-iE_n(k_x)t}w(k_x)|^2$, with $|\psi_{L,n}\rangle$ the left eigenvector. An analytical derivation of $W(k_x,t)$ for edge states is provided in Appendix \ref{app:wpdynamics}.
The real-space edge localization along the cylinder is monitored through $\langle y(t)\rangle=\langle\Psi|\hat y|\Psi\rangle/\langle\Psi|\Psi\rangle$.

For a narrow wave packet at short times, while its width changes negligibly and the dispersion can be linearised over its support, the packet center approximately obeys
$\langle \dot{x} \rangle\simeq\frac{d\,\mathrm{Re}\,E_k}{dk_x}$ and
$\langle \dot{k}_x \rangle\simeq\sigma_k^2\frac{d\,\mathrm{Im}\,E_k}{dk_x}$~\cite{Muschietti1993,dong-2025,He2025}. The former is the usual group-velocity, while the latter describes “self-acceleration” governed by the imaginary dispersion and wave packet width $\sigma_k$. We initialize wave packets at $k_x=\pi$, corresponding to localization at edges $y=0$ or $y=N_y$.
The wave packet dynamics display two central features: 1) The vanishing slope of the real spectrum at $k_x = \pi$ gives the edge mode wave packet zero initial group velocity along the periodic direction, in contrast to the conventional chiral edge modes. 2) The finite slope in the imaginary part of the spectrum $v_I= \left (\partial \mathrm{Im} E)/\partial k_x \right|_{k_x=\pi}$ leads to a local short-time drift of the packet center, $k_c(t)\simeq k_0+v_I\sigma_k^2 t$.  For the edge band, states near $k_x=\pi$ are strongly edge-localized but progressively delocalize toward the EPs (Fig.~\ref{fig:spectrum-pr-ipr}(a),(b)). Thus the drift of the wave packet in $k_x$-space leads to redistribution toward bulk-delocalised states along $y$ with time, making them `ephemeral' edge modes (Fig.~\ref{fig:wavepacket-dynamics-edge} (a)). The fate of the edge mode is decided by the imaginary slope $v_I\sim\tilde{t}\sinh\phi$ that determines the drift. For weak non-Hermiticity ($\phi\ll1$) and narrow wave packets ($\sigma_k\ll1$) \footnote{The relevant comparison scale is set by the momentum-space separation between the EPs, where the imaginary dispersion is finite.}, the drift is strongly suppressed, stabilizing the edge for longer time scales (Fig.~\ref{fig:wavepacket-dynamics-edge}(a)).

To show the finite-time persistence window of edge localisation, we plot the late-time $\langle y(t)\rangle$ as a function of the width, for different values of $\phi$ (Fig.~\ref{fig:wavepacket-dynamics-edge}(b)). Edge localization can persist for wider wave packets when $\phi\ll1$ \footnote{For instance, the small non-reciprocity $\phi=0.05$ used in Fig.~\ref{fig:wavepacket-dynamics-edge} corresponds to $t_2=0.3003$ and $\gamma=0.015$, representing a physically reasonable deviation from the Hermitian limit.}. For larger $\phi$, however, wider wave packets acquire finite self-acceleration and drift away from $k_x=\pi$ towards the maxima of $\mathrm{Im}E_k$, where its slope vanishes. For a single band with a unique accessible maximum, the normalised momentum distribution concentrates asymptotically near that maximum. In the bulk-band example, the selected states are delocalised but weakly polarised away from the strip center as shown in Appendix \ref{app:wpdynamics}. Similar edge-to-bulk redistribution has been reported in recent works~\cite{Zhang2025,jana2025,He2025}.

\section{Conclusion}
Thus, we have investigated the exceptional phase of the imaginary-flux Haldane model, which hosts $\mathcal{PT}$-protected ERs that confine the right-state Berry curvature, together with non-chiral edge states localised at zigzag boundaries on a cylinder. Such a non-chiral edge localisation is in contrast to the chiral modes found in topological phases. These modes are also robust to disorder in the non-reciprocity. The self-acceleration term in the equations of motion leads to redistribution of the edge wave packet into the bulk over time,  thus making them `ephemeral edge modes'. Yet the time scale for edge-to-bulk delocalization can be extended by tuning the non-Hermiticity and the wave packet width. We also account for the observed step-like increase in the number of EP pairs through a zone-folding approximation.

The imaginary-flux Haldane model furnishes a two-dimensional lattice setting for the geometrical, spectral, and dynamical phenomena studied here. While ERs are known to induce nonlinear Hall responses in continuum systems~\cite{Qin2025}, in our case the Berry curvature maintains the same sign across both ERs, unlike in imaginary-mass models, potentially yielding interesting non-linear responses. ERs also emerge in the Majorana-basis Haldane model relevant to the Kitaev honeycomb system in disordered magnetic fields~\cite{bergholtz-2022, fukui-2024}, motivating a separate investigation of possible non-chiral Majorana boundary dynamics.

Our results position the imaginary-flux Haldane model as a natural setting to explore quantum geometric, exceptional-point--driven bulk and edge phenomena, complemented by existing proposals for their realisation in optical traps and topolectrical circuits~\cite{Ezawa2019,dong-2025}. In particular, a recent proposal for generating a uniform imaginary magnetic field for ultracold atoms, using Raman coupling and state-dependent losses, also predicts width-dependent wave-packet transport~\cite{alminas2026}, suggesting that the amplification-driven dynamics discussed here may be accessible in cold-atom settings. Another natural extension is to characterise the singular exceptional structure of this limit within the broader parameter space of the complex-flux Haldane model.
\begin{acknowledgments}
SH would like to acknowledge funding support from the ANRF Prime Minister Early Career Grant (PM-ECRG): ANRF/ECRG/2024/002083/PMS. AP acknowledges the Institute Fellowship from the Indian Institute of Science Education and Research, Thiruvananthapuram (IISER TVM). Both AP and HC acknowledge institutional support and research facilities provided by IISER TVM.
\end{acknowledgments}

% --- Start Appendix ---
\appendix
\section{Imaginary flux Haldane model}
\label{app:main-model}

{\it Transfer Matrix analysis:}
\label{app:transfer-matrix}
We analyze the edge mode wavefunction and spectrum using the transfer-matrix method. The formalism employed here closely follows the general framework developed in ~\cite{mizoguchi-2021}, which we adapt to the present non-Hermitian model.  Under zigzag boundary conditions the Hamiltonian is given by Eq.~\eqref{eq:obc-hamiltonian}. By writing the edge mode as $\Psi(k_x)_{\text{Edge}}^{\dagger}= \sum_i^{N_y} \psi_i(k_x)^{\dagger} \phi_i$, the equation of motion gives the following recurrence relation:

 \begin{equation}
A_+\phi_{i+1} + A_-\phi_{i-1} + B\phi_i = E(k_x)\phi_i,
\label{eq:recurrence}
\end{equation}
where $\phi_i$ denotes the  two-component amplitude in cell $i$, $h_{i,i}=B(k_x)$ and $h_{i,i\pm1}=A_\pm(k_x)$. In general $A_-\ne A_+^\dagger$.
When $A_+$ is invertible, the transfer matrix for the above is given by
\begin{align}
\begin{pmatrix}
\phi_{i+1}(k_x) \\
\phi_i(k_x)
\end{pmatrix}
=
T(k_x)\,
\begin{pmatrix}
\phi_i(k_x) \\
\phi_{i-1}(k_x)
\end{pmatrix}.
\end{align}

\begin{figure*}[t!]
\centering\normalcolor

\begin{minipage}[t]{0.24\textwidth}
    \centering
    \includegraphics[width=\linewidth]{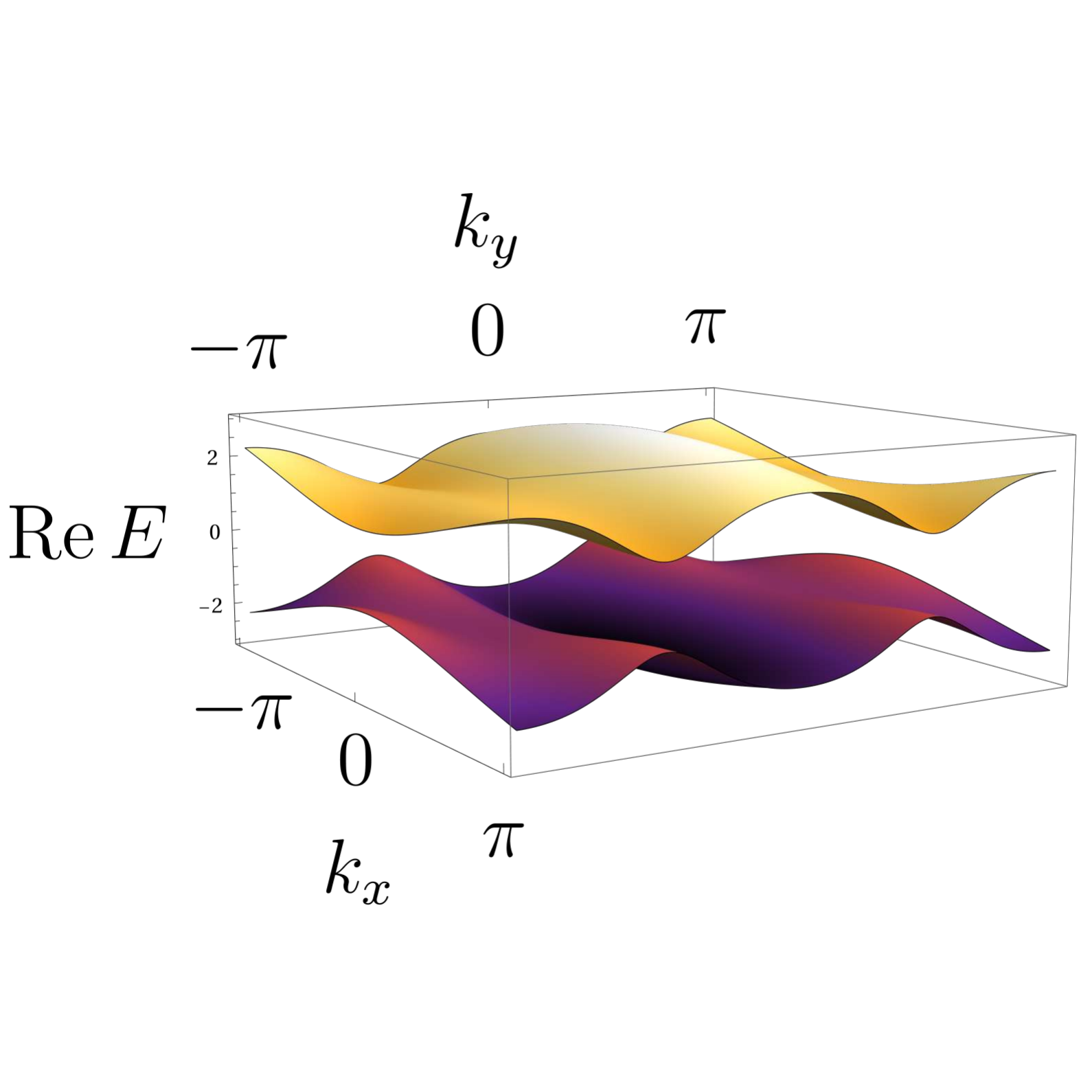}
    \\ (a)
\end{minipage}
\hfill
\begin{minipage}[t]{0.24\textwidth}
    \centering
    \includegraphics[width=\linewidth]{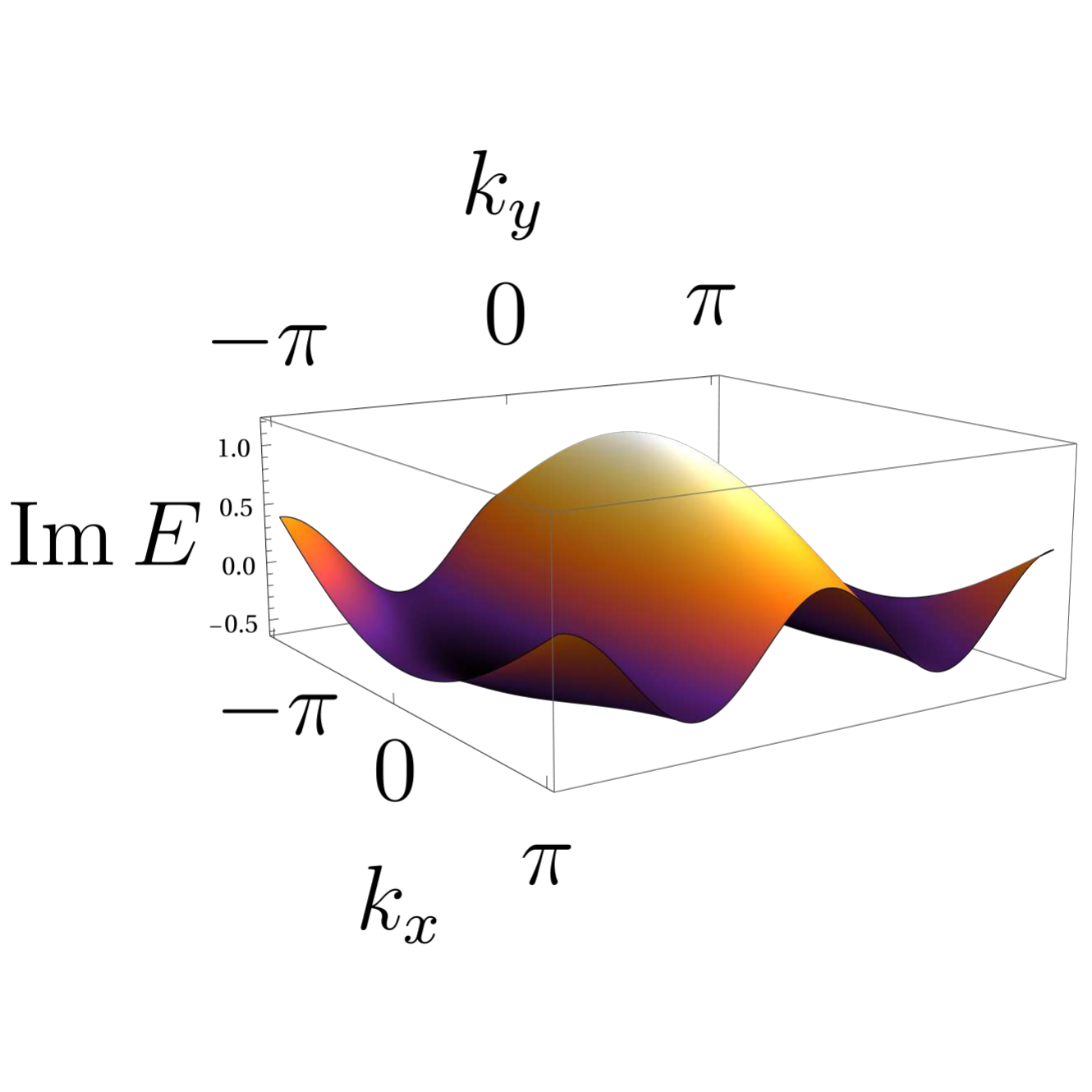}
    \\ (b)
\end{minipage}
\hfill
\begin{minipage}[t]{0.24\textwidth}
    \centering
    \includegraphics[width=\linewidth]{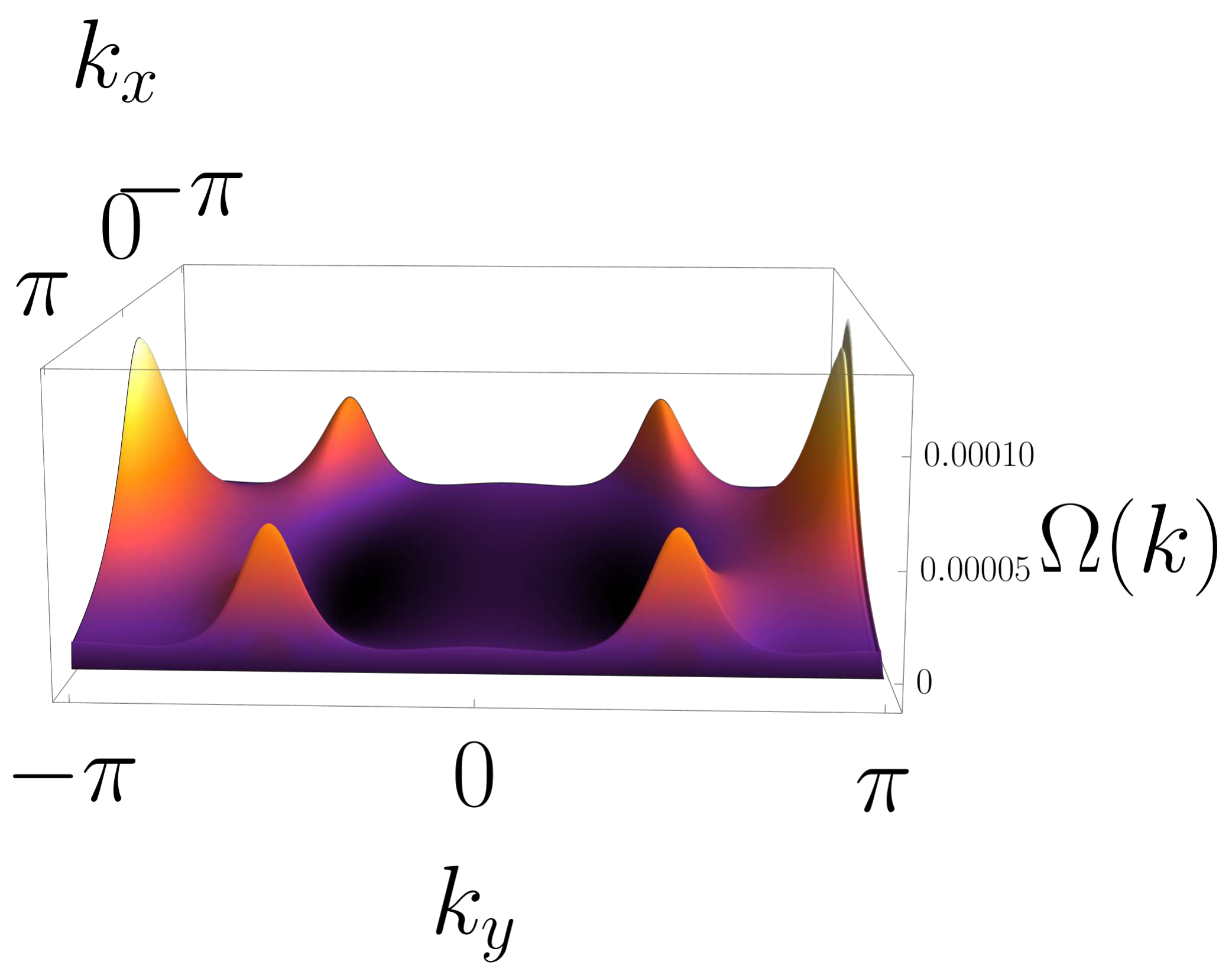}
    \\ (c)
\end{minipage}
\hfill
\begin{minipage}[t]{0.24\textwidth}
    \centering
    \includegraphics[width=\linewidth]{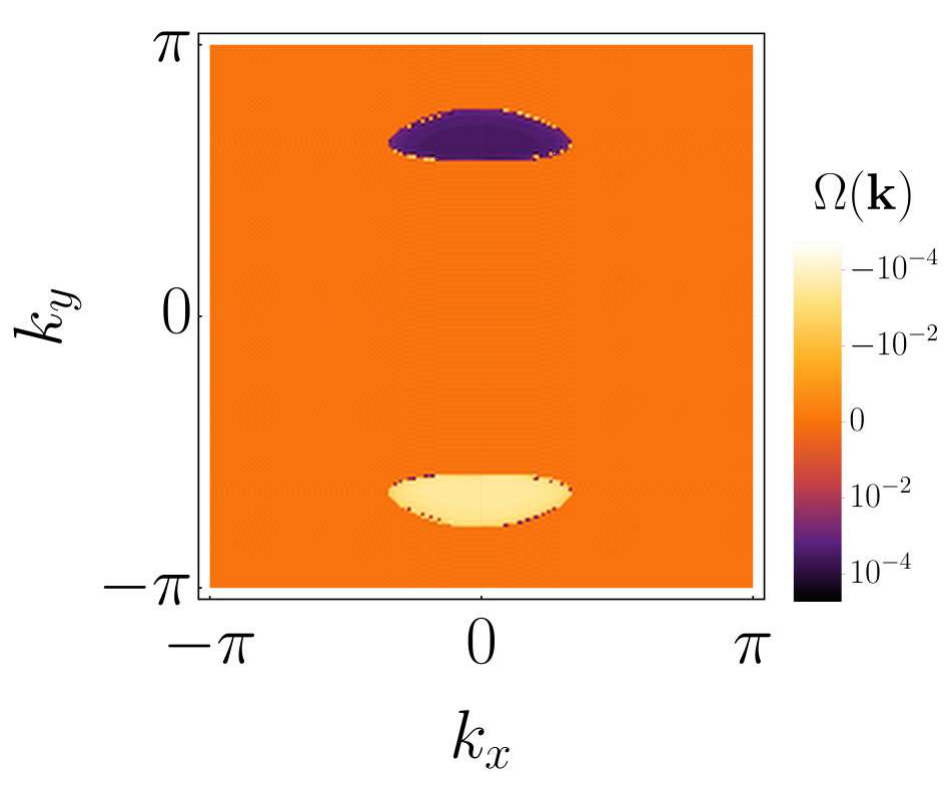}
    \\ (d)
\end{minipage}

\caption{\justifying
(a),(b) Real and imaginary parts of the energy spectrum of the complex-flux Haldane model under periodic boundary conditions. Away from the purely imaginary-flux case, this model does not exhibit any exceptional points.
(c) Berry curvature distribution for the right eigenstates of the complex-flux Haldane model, which is smoothly distributed over the Brillouin zone. (Parameters for (a)-(c) : $t_1 = 1$, $\tilde{t} = 0.22$, $\phi = 0.1$, $\Phi=\frac{\pi}{2}$.) (d) Berry curvature distribution for graphene with an imaginary Semenoff mass. The curvature is concentrated within the exceptional rings and exhibits opposite signs between the two rings. (Parameters for (d) : $t_1 = 1$, $M_A=-M_B=0.5i$.)}
\label{fig:complex-flux-all}
\end{figure*}
\begin{figure}[t]
\centering\normalcolor
    \includegraphics[width=6.54cm, height=4cm]{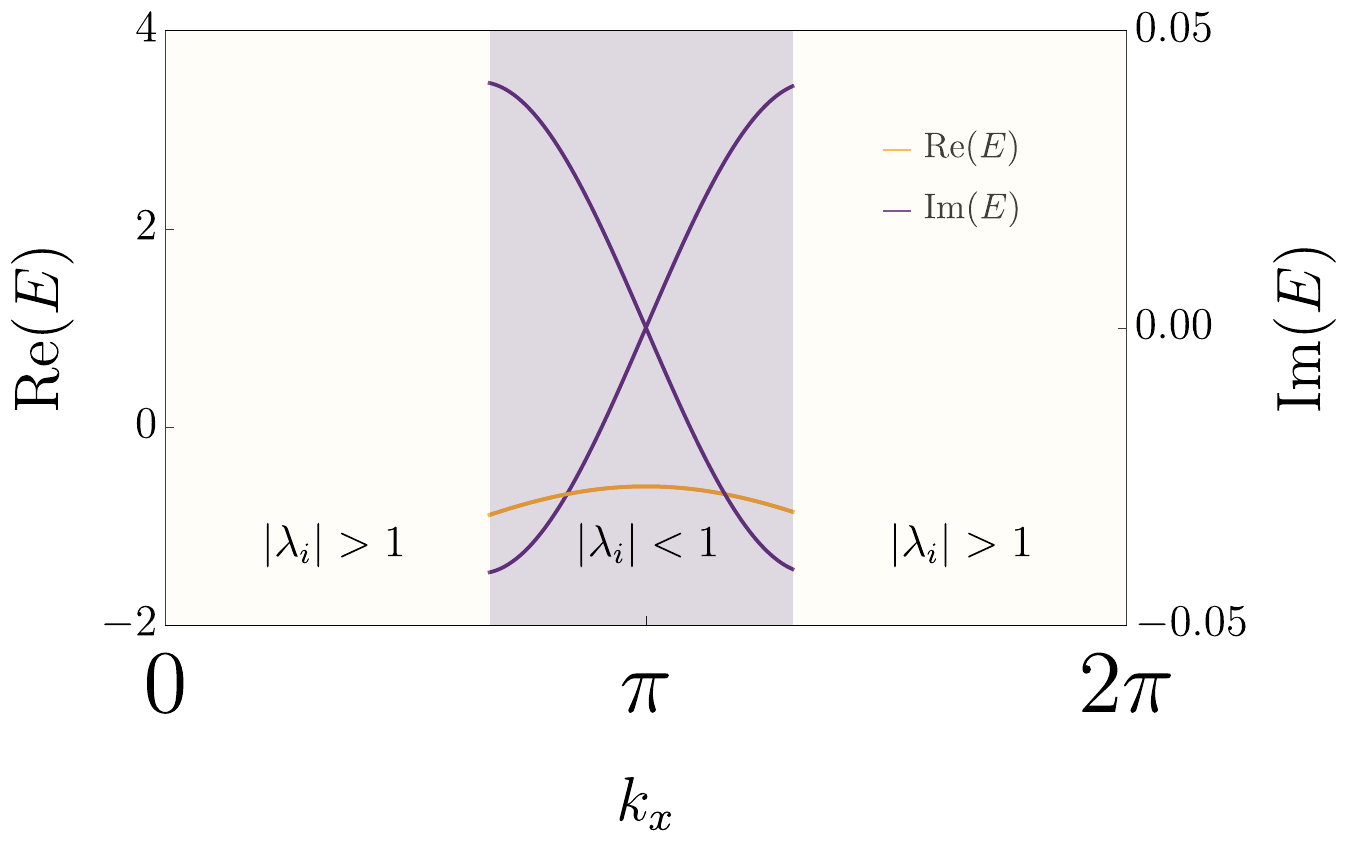}
\caption{\justifying
Behaviour of the eigenvalues $\{\lambda_i\}$ of the transfer matrix.
A bifurcation transition from edge localisation to bulk delocalisation occurs when $|\lambda_i|=1$. (Parameters: $N_y = 30$, $t_1 = 1$, $\tilde{t} = 0.3$, $\phi = 0.1$)}
\label{fig:lyapunov}
\end{figure}

{\small\setlength{\arraycolsep}{2pt}\begin{equation*}
T(k_x)=
\begin{pmatrix}
A_+^{-1}(k_x)\big[E(k_x)I_2 - B(k_x)\big] & -A_+^{-1}(k_x)A_-(k_x) \\
I_2 & 0
\end{pmatrix}
\end{equation*}}
Solving for the eigenvalues $\{ \lambda_i\}$ of the transfer matrix \cite{vatsal-2016,kawabata-2018}, we obtain the edge mode spectrum and the parameter range for their existence by combining decaying roots $|\lambda_i|<1$ and imposing the boundary equations. The decay length $\xi=-1/\ln|\lambda|$ diverges at $|\lambda|=1$ in the semi-infinite limit as shown in the Fig. \ref{fig:lyapunov}.

 {\it Winding number:}
There is a winding number associated with the edge mode appearing at $k_x=\pi$. One can reinterpret the Hamiltonian in Eq.~\eqref{eq:obc-hamiltonian} as a family of 1D Hamiltonians with $k_x$ as an external parameter. Using the framework used to characterise the edge states of the Haldane model \cite{mahmood-2025},  our model can be rewritten as a 1D effective model for each  value of $k_x$ (imposing PBC along $y$):
\begin{equation}
H(k_y) = d_0(k_y)\,\sigma_0 +d_x(k_y)\,\sigma_x +d_y(k_y)\,\sigma_y + d_z(k_y)\,\sigma_z
\label{eq:Hky}
\end{equation}
where the components of $\mathbf d(k_y)$ are
$d_0(k_y)=2\tilde{t}\cosh\phi[\cos k_x+2\cos(\tfrac{k_x}{2})\cos k_y], d_x(k_y)=t_1[2\cos(\tfrac{k_x}{2})+\cos k_y], d_y(k_y)=-t_1\sin k_y,$ and $d_z(k_y)=2i\tilde{t}\sinh\phi[-\sin k_x+2\sin(\tfrac{k_x}{2})\cos k_y].$
At $k_x=\pi$, set $C=\tilde t\cosh\phi$ and $b=4\tilde t\sinh\phi$. The shifted Hamiltonian is
\begin{equation}
H_\pi+2C\sigma_0=(t_1\sigma_x+ib\sigma_z)\cos k_y-t_1\sigma_y\sin k_y.
\end{equation}
For $t_1>0$ and $|b|<t_1$, the momentum-independent similarity transformation $V=e^{\eta\sigma_y/2}$, with $\tanh\eta=b/t_1$, gives
\begin{equation}
V(H_\pi+2C\sigma_0)V^{-1}=a\cos k_y\,\sigma_x-t_1\sin k_y\,\sigma_y,
\end{equation}
where $a=\sqrt{t_1^2-b^2}$. The transformed Hamiltonian is Hermitian and anticommutes with $\sigma_z$. Its winding is
\begin{equation}
\nu=\frac{1}{2\pi}\int_{-\pi}^{\pi}dk_y\,\partial_{k_y}\arg(a\cos k_y-it_1\sin k_y)=-1.
\end{equation}
The spectrum satisfies
$(E+2C)^2=t_1^2-b^2\cos^2 k_y$, which implies that the gap closes at $|b|=t_1$, and marks the threshold for defining a valid winding number. Away from $k_x=\pi$, this construction does not supply a quantised invariant, where other characterizations of edge-modes would be useful~\cite{Jiang2018,Hattori_2024,Zhong2025}. The disorder calculation in the main text demonstrates persistence of the mode near $\mathrm{Im}E=0$ in the parameter range studied.
\section{Complex-flux Haldane model}
\label{app:complexflux}

In this section, we extend the construction of the non-Hermitian Haldane model further by promoting the flux to be a fully complex quantity $\Phi+i\phi$:
\begin{displaymath}{
\begin{aligned}
\mathcal{H}_{\mathcal{CFH}} =\;&
\sum_{i}  M_i c_i^\dagger c_i
+ t_1 \sum_{\langle i,j\rangle} c_i^\dagger c_j \\
&+ \tilde{t} \sum_{\langle\!\langle i,j\rangle\!\rangle}
\Big[
e^{i(\Phi+i\phi)} c_i^\dagger c_j
+ e^{-i(\Phi+i\phi)} c_j^\dagger c_i
\Big]
\end{aligned}
}\end{displaymath}
 For real $M_i$, the model reduces to the Hermitian Haldane model at $\phi=0$ and $\Phi=\frac{\pi}{2}$. Imposing periodic boundary conditions along both lattice directions allows us to express the Hamiltonian in momentum space, but with a complex flux.

We note that the time-reversal symmetry is preserved at $\Phi=0$ modulo $\pi$ for real on-site masses and is broken at other values of the real flux. With broken TRS, in some parameter regimes the model exhibits an isolated band with a nonzero Chern number separated from the other band by a well-defined gap (Figs.~\ref{fig:complex-flux-all}(a),(b)). There are chiral edge states on imposing open boundary conditions but with finite lifetimes{~\cite{dong-2025}}. We additionally compute the Berry curvature of this model and find that its distribution differs remarkably from that obtained in our system in that it is smoothly distributed throughout the BZ in contrast with the confinement in ERs (See Fig.~\ref{fig:complex-flux-all}(c)).

We also consider a model with an imaginary Semenoff mass term ($M_A=-M_B=i\Gamma$, with real $\Gamma$) and no NNN hoppings. Although this model also hosts ERs, its Berry curvature landscape differs from our model. In particular, one can observe from Fig.~\ref{fig:complex-flux-all}(d) that the Berry curvature has opposite-signed contributions between the two ERs across the BZ.

\section{Wave packet distribution in $k_x$ space and dynamics of bulk states}
\label{app:wpdynamics}
We construct a wave packet from a subset of bands, weighted by a Gaussian
$w(k_x)=\frac{1}{\sqrt{2\pi\sigma_k^2}}\exp[-(k_x-k_0)^2/(2\sigma_k^2)]$.
The time-evolved wave-packet is given by
$\Psi(y,t)=\sum_m \int dk_x\, e^{-iE_m(k_x)t}\, w(k_x)\, |\psi_{R,m}(k_x,y)\rangle$,
where $|\psi_{R,m}(k_x,y)\rangle$ is the $m$-th right eigenvector.  For an isolated band, $W(k_x,t)=|\langle \psi_{L,n}(k_x)|\Psi(y,t)\rangle|^2
=|e^{-iE_n(k_x)t}w(k_x)|^2$, with $|\psi_{L,n}\rangle$ the left eigenvector.

Focusing on the edge band $E_m=E_{EB}$ and using
$|e^{-iE_{EB}t}|^2=e^{2\,\mathrm{Im}E_{EB}(k_x)t}$, we obtain
\begin{equation}
W(k_x,t)
\propto
\exp\!\left[
-\frac{(k_x-k_0)^2}{\sigma_k^2}
+2\,\mathrm{Im}E_{EB}(k_x)\,t
\right].
\label{eq:W_general}
\end{equation}

\begin{figure}
    \centering
    \includegraphics[width=1\linewidth]{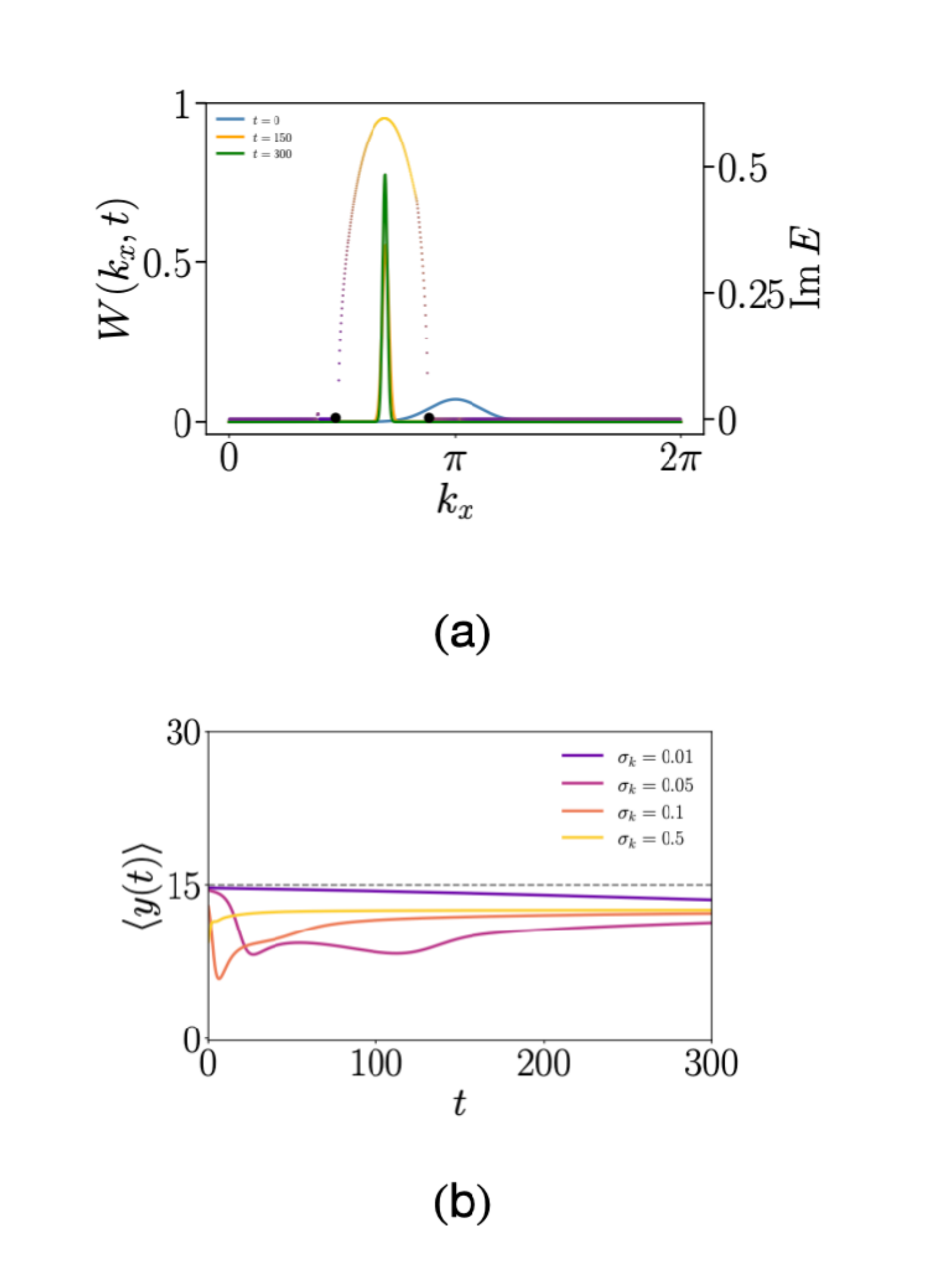}
    \caption{\justifying
    (a) Wave packet distribution dynamics for a bulk band hosting two EPs.
    An initial wave packet prepared around an EP concentrates near the momentum corresponding to the largest accessible value of the imaginary part of the spectrum.
    (b) Expectation value $\langle y(t) \rangle$ for different initial momentum widths over a bulk band.
    The late-time value of $\langle y(t) \rangle$ shifts away from the center of the system,
    indicating the polarized nature of the states at the maximum of the imaginary spectrum.
    (Parameters: $N_y = 30$, $\tilde{t} = 0.3$, and $\phi = 0.45$.)}

    \label{fig:bulk-dynamics}
\end{figure}

To determine the dynamics near the packet center $k_0$, we expand the imaginary part of the dispersion up to second order: $\mathrm{Im}E_{EB}(k_x)
= \mathrm{Im}E_{EB}(k_0) + v_I (k_x-k_0) + \tfrac{1}{2} u_I (k_x-k_0)^2$,
where $v_I=\left.\partial_{k_x}\mathrm{Im}E_{EB}\right|_{k_0}$
and $u_I=\left.\partial_{k_x}^2\mathrm{Im}E_{EB}\right|_{k_0}$. Substituting this expansion into Eq.~\eqref{eq:W_general} and collecting quadratic terms gives $W(k_x,t)\propto\exp\!\left[-\big(\tfrac{1}{\sigma_k^2}-u_I t\big)(k_x-k_0)^2 +2 v_I t (k_x-k_0)\right]$. Completing the square yields a Gaussian form:  $W(k_x,t)\propto\exp\!\left[\frac{v_I^2 t^2}{\tfrac{1}{\sigma_k^2}-u_I t}\right]\exp\!\left[-\frac{(k_x-k_c(t))^2}{\tilde{\sigma}_k^2(t)}\right]$  where the center and width evolve as
\begin{equation}k_c(t)
= k_0 + v_I t\,\tilde{\sigma}_k^{\,2}(t),
\quad \tilde{\sigma}_k^{\,2}(t)=
\frac{1}{\tfrac{1}{\sigma_k^2}-u_I t}
\label{eq:center_width}
\end{equation}
This local approximation requires $1-u_I\sigma_k^2t>0$ and packet support within the Taylor expansion region. Thus, for short times such that $|u_I|\sigma_k^2 t\ll1$, the time-dependent width
$\tilde{\sigma}_k^{\,2}(t)=\sigma_k^2/(1-u_I\sigma_k^2 t)\approx
\sigma_k^2(1+u_I\sigma_k^2 t)$, so that the center evolves as
$k_c(t)=k_0+v_I t\,\tilde{\sigma}_k^{\,2}(t)\approx
k_0+v_I\sigma_k^2 t+\mathcal{O}(t^2)$.
Thus the packet drift is controlled by the slope
$\partial_{k_x}\mathrm{Im}E$, while its deformation is governed by the curvature
$\partial_{k_x}^2\mathrm{Im}E$. In particular, $u_I>0$ leads to broadening, $u_I<0$ to squeezing, and $u_I=0$
implies pure drift without width change to leading order.

%\sout{The fixed-width drift law $\dot k_c\simeq v_I\sigma_k^2$ holds only while the packet is narrow and its width changes negligibly. The covariance equation above remains the single-band result beyond this local approximation. For a bulk band between EPs, differential amplification favors the largest accessible $\mathrm{Im}E$ on the initially populated support. If that maximum is unique, $P$ concentrates near it asymptotically [Fig.~\ref{fig:bulk-dynamics}(a)]. }%A local maximum need not dominate globally, and there is no generic finite time at which the distribution stops evolving. Degenerate maxima can retain contributions from more than one momentum.

% The normalised transverse center is
% \begin{equation}
% \langle y(t)\rangle=\int dk_x\,P(k_x,t)\langle\psi_R(k_x)|\hat y|\psi_R(k_x)\rangle.
% \end{equation}
We also study the dynamics of the bulk bands hosting EPs. As in the case of the edge band, there exists a maximum in the imaginary part of the dispersion for these states. For a single band with a unique accessible maximum, the normalised momentum distribution concentrates asymptotically near that maximum (See Fig.~\ref{fig:bulk-dynamics}(a))

The states around the maxima are special from the view of their localisation properties along $y$-direction.  These states are delocalised into the bulk along $y$ but have a small polarisation away from the center  as illustrated in Fig.~\ref{fig:bulk-dynamics}(b).   Therefore, ``center of mass'' $\langle y(t)\rangle$ converges to a steady state value that is shifted away from the center of the system.

\section{Zone-folding description of the EP cascade}
\label{app:SM-zone-folding}

The multiple bulk EP pairs in the cylinder geometry with zigzag edges (Fig.~\ref{fig:edge-exceptional-band}(a) of the main text) can be estimated in a bulk-channel approximation from the ERs in the spectrum under full periodic boundary conditions through zone-folding. They originate from the discrete standing-wave modes that arise in the zigzag cylinder. The momentum along the direction of periodicity is conserved. Open boundaries in the transverse direction break transverse
translation symmetry, so $k_y$ is no longer a conserved
Bloch momentum. Nevertheless, the extended bulk-like states
form a discrete set of transverse standing-wave modes, which
may be parametrised by wave numbers $q_{\perp,n}$. The subscript $n$ labels the discrete values of $q_{\perp}$. In this approximation each $q_{\perp,n}$ defines a one-dimensional bulk subband described by Eq.~\eqref{eq:folded-H}. Boundary phase shifts and mode mixing prevent an exact finite-strip decomposition into independent channels. Since the subsequent calculations are carried out for a fixed $n$, we drop this subscript hereafter for simplicity.

\subsection{Two-dimensional bulk Hamiltonian in zigzag coordinates}
We introduce $k_x\equiv\theta_1-\theta_2$ and $q_\perp=(\theta_1+\theta_2)/2$, i.e.\ $\theta_{1,2}=q_\perp\pm k_x/2$. Sum-to-product identities then bring the Bloch Hamiltonian of the main text to the form
\begin{equation}
H(k_x,q_\perp)=d_0\sigma_0+d_x\sigma_x+d_y\sigma_y+i\Gamma\sigma_z,
\label{eq:folded-H}
\end{equation}
with $d_0=2\tilde t\cosh\phi\,[\cos k_x+2c\cos q_\perp]$, $d_x+id_y=t_1[1+2c\,e^{iq_\perp}]$ and $\Gamma=4\tilde t\sinh\phi\,s\,(\cos q_\perp-c)$. Here $c\equiv\cos(k_x/2)$, $s\equiv\sin(k_x/2)$, and $h_3=i\Gamma$. Fixing $q_\perp=q_{\perp,n}$ gives a two-band Hamiltonian in $k_x$ for each transverse channel.

\subsection{Zone-folded bulk EP positions}
The term $d_0$ shifts both levels equally, so $E_\pm=d_0\pm\sqrt{\Delta}$. The discriminant is
$\Delta=d_x^2+d_y^2-\Gamma^2=t_1^2(1+4c^2+4c\cos q_\perp)-16\tilde t^{\,2}\sinh^2\!\phi\,s^2(c-\cos q_\perp)^2$,
and generic EPs occur at $\Delta=0$ with $H-d_0\sigma_0\ne0$. Read at fixed $k_x$, this condition gives the $\mathcal{PT}$-transition curve of each channel:
$\sinh\phi_c(k_x)=|t_1|\sqrt{1+4c^2+4c\cos q_\perp}\,/\,\big(4|\tilde t|\,s\,|c-\cos q_\perp|\big)$.
This expression assumes a nonzero denominator; otherwise the original discriminant must be used. For a given $\phi$, the periodic-bulk EPs are the intersections of the line $\phi=\mathrm{const.}$ with this curve.

To find the EP momenta explicitly, we use $s^2=1-c^2$ and define $\kappa=4\tilde t\sinh\phi/t_1$. The condition $\Delta=0$ then becomes a quartic in $c$:
\begin{equation}
\begin{aligned}
0={}&\kappa^2c^4-2\kappa^2\cos q_\perp\,c^3+(4-\kappa^2\sin^2q_\perp)\,c^2\\
&+(4+2\kappa^2)\cos q_\perp\,c+1-\kappa^2\cos^2q_\perp .
\end{aligned}
\label{eq:EP-quartic}
\end{equation}
The map $k_x=2\arccos c$ is one-to-one between $[0,2\pi]$ and $[-1,1]$. Each real root $c\in[-1,1]$ therefore fixes one EP momentum, giving at most four EPs per channel. A root is a genuine EP only if it also satisfies $H-d_0\sigma_0\neq0$, i.e.\ $d_x$, $d_y$ and $\Gamma$ do not vanish simultaneously. Eq.~\eqref{eq:EP-quartic} is exact for the periodic bulk. Applying it to the strip inherits the channel approximation.
\begin{figure}
    \centering\normalcolor
    \includegraphics[width=0.75\linewidth]{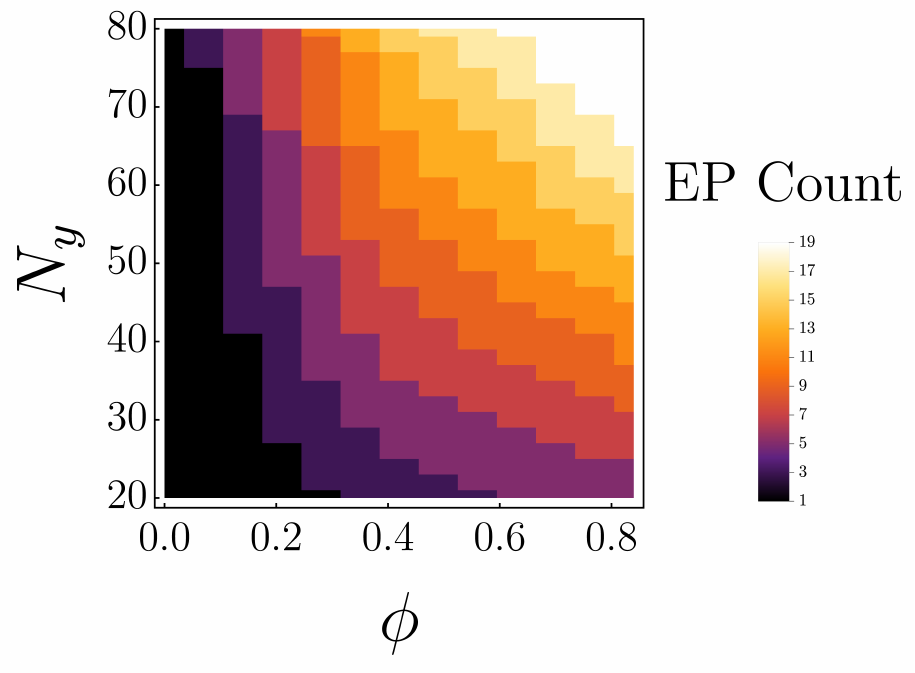}
    \caption{\justifying Total number of EP pairs (color scale) in the zigzag cylinder versus non-reciprocity $\phi$ and width $N_y$. For fixed low-lying channels counted from a valley, the onset shifts to smaller $\phi$ approximately as $N_y^{-1}$, producing the staircase of Eq.~\eqref{eq:EP-count}.}
    \label{fig:width-dependence}
\end{figure}
\textit{EP order:} Eq.~\eqref{eq:folded-H} obeys $(H-d_0\sigma_0)^2=\Delta\,\sigma_0$. At $\Delta=0$ with $H-d_0\sigma_0\neq0$, the matrix $H-d_0\sigma_0$ is nonzero and nilpotent, so it has a single eigenvector. Hence generic solutions are EP2s. Where $\partial_{k_x}\Delta\neq0$, the splitting behaves as $E_+-E_-\propto\sqrt{k_x-k_x^{\rm EP}}$.  This two-band argument does not exclude fine-tuned multilevel degeneracies of the full strip.

\textit{Weak non-reciprocity:} At $\phi=0$, the conditions $d_x=d_y=0$ give the Dirac points $K=(2\pi/3,\pi)$ and $K'=(4\pi/3,0)$. Since $\Gamma\propto\sinh\phi$, the weak-$\phi$ EPs grow out of these points. Near $K$ we set $k_x=2\pi/3+\delta k_x$ and $q_\perp=\pi+\delta q_\perp$. To leading order, $d_x^2+d_y^2\simeq t_1^2(\tfrac34\delta k_x^2+\delta q_\perp^2)$ and $\Gamma^2\simeq27\tilde t^{\,2}\sinh^2\!\phi$. The ER is therefore the ellipse $\tfrac34\delta k_x^2+\delta q_\perp^2=27(\tilde t/t_1)^2\sinh^2\!\phi$. A channel at distance $\delta q_\perp$ cuts this ellipse at
\begin{equation}
k_x^{\mathrm{EP}\pm}\simeq\frac{2\pi}{3}\pm\frac{2}{\sqrt3}\sqrt{27\Big(\frac{\tilde t}{t_1}\Big)^{\!2}\sinh^2\!\phi-\delta q_\perp^{2}},
\label{eq:qx-Dirac}
\end{equation}
and analogously near $K'$. Channels close to $q_{\perp D}$ cut the ellipse near its widest chord and give the outer pairs. More distant channels cut near its tips and give the inner pairs.

\subsection{Width dependence and the step structure}
For large $N_y$, the open boundaries select the transverse wave numbers $q_{\perp,n}=\pi(n+\eta_n)/N_y+O(N_y^{-2})$~\cite{BreyFertig2006}. The boundary phase shift $\eta_n=O(1)$ can depend on $k_x$, $E$ and the couplings. It alters threshold prefactors but not the scaling. A mode contributes EPs once the radicand in Eq.~\eqref{eq:qx-Dirac} becomes non-negative, i.e.\ for $\sinh\phi_{c,n}\simeq|t_1||\delta q_{\perp,n}|/(3\sqrt3|\tilde t|)$. For a fixed low-lying mode counted from the valley, $|\delta q_{\perp,n}|=O(N_y^{-1})$. Because $\phi_{c,n}\simeq\sinh\phi_{c,n}$ at weak non-reciprocity, this gives $\phi_{c,n}\propto N_y^{-1}$ (Fig.~\ref{fig:width-dependence}).

Each crossed threshold adds one EP pair at each valley. In the observed staircase regime, the total number of EPs, $N_{\rm EP}$, is summarised schematically by
\begin{equation}
\frac{N_{\mathrm{EP}}}{2}=1+2\sum_n\Theta(\phi-\phi_{c,n}),
\label{eq:EP-count}
\end{equation}
where $n$ runs over valley-related channel pairs and the leading $1$ is the edge--bulk pair. At fixed $\phi$, the number of contributing modes grows roughly linearly with $N_y$ until finite-band saturation sets in. At larger $\phi$ the EPs leave the Dirac neighbourhoods, and their approximate positions follow from Eq.~\eqref{eq:EP-quartic}. A finite cylinder has finitely many modes, so the staircase eventually saturates or undergoes root rearrangements.

\textit{Edge--bulk pair.} The outermost pair lies outside the zone-folding picture, which assumes a real $q_\perp$. Edge states are instead characterised by the transfer-matrix decay factor, with $|\lambda|<1$ (Sec.~\ref{app:main-model}). The limit $|\lambda|\to1$ marks the loss of exponential localization in the semi-infinite limit. An edge--bulk EP additionally requires a defective finite-strip eigenvalue, and a finite-width hybridization may shift it slightly from that boundary. Numerically, it coincides with vanishing phase rigidity and suppressed IPR (See Fig.~\ref{fig:spectrum-pr-ipr}(c)).

%\nocite{*}
% \bibliographystyle{apsrev4-2}
\bibliography{references}

\end{document}